\documentclass[a4paper,11pt]{article}
\pdfoutput=1 % if your are submitting a pdflatex (i.e. if you have images in pdf, png or jpg format)

\usepackage{jheppub}

\usepackage[ddmmyyyy,hhmmss]{datetime}

\usepackage{amsmath}
\usepackage{amsfonts}
\usepackage[retainorgcmds]{IEEEtrantools}
\usepackage{tikz-feynman}
\usepackage{float}
\usepackage{verbatim}
\usepackage{slashed}
\usepackage{empheq}
\usepackage[normalem]{ulem}

\hypersetup{pdfstartview={XYZ null null 0.75}}
\allowdisplaybreaks

\makeatletter \@addtoreset{equation}{section} \makeatother

\newcommand{\OO}{\mathcal{O}}
\newcommand{\LL}{\mathcal{L}}

\newcommand{\I}{\mathcal{I}}

\newcommand{\Y}{\mathbb{Y}}
\newcommand{\N}{\mathcal{N}}

\newcommand{\G}{\mathcal{G}}

\newcommand{\K}{\mathcal{K}}

\usepackage{amsmath,graphicx,amssymb,subfigure,bbm,comment}
\usepackage[all]{xy}
\usepackage{mathtools}
\usepackage{amsfonts}
\usepackage{comment}
\usepackage{mdframed}

\usepackage{feynmp}
\tikzfeynmanset{compat=1.1.0}

\usepackage{xcolor}
\usepackage{pifont}

\allowdisplaybreaks

\def\be{\begin{equation}}
\def\ee{\end{equation}}
\def\ba{\begin{eqnarray}}
\def\ea{\end{eqnarray}}

\def\a{\alpha}

\def\b{\beta}

\def\g{\gamma}
\def\G{\Gamma}
\def\d{\delta}
\def\D{\Delta}

\def\l{\lambda}
\def\L{\Lambda}
\def\s{\sigma} 
\def\S{\Sigma}

\def\cN{{\cal N}}

\def\cL{{\cal L}}

\def\IR{\relax{\rm I\kern-.18em R}}

\def\IR{\relax{\rm I\kern-.18em R}}
\def\inv{^{\raise.15ex\hbox{${\scriptscriptstyle -}$}\kern-.05em 1}}

\def\cL{{\cal L}}

\title{%Interpolating holographic defect CFTs\\
Holographic interpolations of defect CFTs}

\author{George Georgiou}
\author[a]{and Dimitrios Zoakos}

\affiliation[a]{Department of Physics, University of Patras, 26504 Patras, Greece.}

\emailAdd{georgios.georgiou2@gmail.com}
\emailAdd{dzoakos@upatras.gr}

\abstract{
We propose a new class of holographic dualities between certain, generically non supersymmetric, 
defect conformal field theories (dCFTs)
and their gravity duals. Our construction interpolates between the 1/2-BPS D3-D3 system and its field theory dual at one end, and the holographic duality presented in \cite{Georgiou:2025mgg} at the other. On the gravity side, the defect is realised by a novel D5 probe brane embedded in the $AdS_5\times S^5$ geometry. The symmetry of the induced on the D5 brane metric is $AdS_3\times S^1\times S^2$. At a certain limit the D5 brane becomes singular and resembles the D3-D3 system.
Consistency requires the presence of two D7 branes on which the D5 brane terminates.
The existence of boundaries induces
a gauge anomaly for the D5 brane which is cancelled through anomaly inflow from the D7
branes. The full system of the D5 and D7 branes is, thus, anomaly free. Also it does not have any tachyonic instabilities for a certain range of its parameters. On the field theory side, we determined the classical solution of the ${\cal N}=4$ SYM equations of motion which
we conjecture to describe the  defect  dual to the D5-D7 system and comment on the identification of the parameters appearing at the two sides of the duality.}

\begin{document}
\maketitle
\flushbottom

%%%%%%%%%%%%%%%%%%%%%%%%%%%%%%%%%%%%%%%%%%%%%%%%%%%%%%%

\section{Introduction}
%The dynamics of conformal field theories (CFTs) in the presence of defects describes a plethora of physical systems, ranging from boundaries and interfaces to objects such as Wilson loops, strings and branes. In general, the presence of a defect breaks, partially or even completely, the conformal symmetry of the ambient CFT rendering the calculation of observables much more involved. The situation becomes even more intricate if one is interested in the behaviour of the theory in the strong coupling regime. The calculations become rather tractable in the case where the ambient CFT is the maximally supersymmetric gauge theory in 4 dimensions, i.e ${\cal N}=4$ SYM. The reason is two-fold. On one hand, the ambient CFT is believed to be integrable in the planar limit \cite{Minahan:2002ve,Bena:2003wd} and on the other because of its holographic description in terms of type IIB string theory on $AdS_5\times S^5$ \cite{Maldacena:1997re}.

Surface operators are extended defects supported on submanifolds of spacetime. They play an essential role in modern gauge theory, particularly in the study of dualities and non-perturbative dynamics. In the context of four-dimensional supersymmetric field theories, co-dimension 2 surface operators, operators supported on two-dimensional surfaces, provide a natural generalization of Wilson and 't Hooft line operators, encoding singular boundary conditions for fields along the two-dimensional defect.

In ${\cal N}=4$  supersymmetric Yang-Mills (SYM) theory, co-dimension 2 surface operators may preserve a fraction of the parent supersymmetry (SUSY) or  break SUSY completely \cite{Georgiou:2025mgg}. They are particularly interesting due to the highly constrained structure of the theory and its exact S-duality symmetry. These operators are typically characterized by prescribing singular behaviour of the gauge and scalar fields near the support of the defect \cite{Gukov:2006jk,Gaiotto:2008sa}. They can be engineered by coupling the 4d theory to a 2d sigma model on the defect or realised via dimensional reduction from the six-dimensional 
(2,0) theory \cite{Kapustin:2006pk,Witten:2009at}.

Beyond their intrinsic interest within quantum field theory, surface operators in 
${\cal N}=4$  SYM have deep connections with geometric representation theory and the geometric Langlands programme. In particular, they appear in the work of Gukov and Witten as parameters specifying ramification in the geometric Langlands correspondence \cite{Gukov:2006jk,Gaiotto:2009hg}. Under S-duality, surface operators transform in non-trivial ways, often involving a duality between strong and weak coupling descriptions and between electric and magnetic data  \cite{Gukov:2006jk}.

In the AdS/CFT correspondence, surface operators correspond to extended objects in the bulk string theory, often realised as probe branes wrapping appropriate submanifolds of the $AdS_5\times S^5$ background. For co-dimension 2 defects in the boundary theory, the natural holographic duals are brane configurations whose worldvolume contains an $AdS_3$  factor, encoding the conformal symmetry preserved by the defect, and wraps internal cycles in the compact space. These branes can backreact to generate full supergravity solutions with localized $AdS_3$ geometries, or can be studied in the probe approximation when the defect is heavy but sparse.
A canonical example, albeit of co-dimension 1, is the D3-D5 brane intersection \cite{Karch:2000gx}, where the defect SYM is realized by a stack of D3-branes ending on a D5-brane. The corresponding defect in 
${\cal N}=4$  SYM preserves 
N=(4,4) supersymmetry. Another example is the 1/2-BPS D3-D3 brane intersection whose gravity dual is a configuration of a probe D3-brane wrapping $AdS_3\times S^1$  inside $AdS_5\times S^5$  \cite{Drukker:2008wr}. The backreacted geometry for this system is known and can be found in \cite{Gomis:2007fi}. Similarly, more general surface operators with reduced supersymmetry or additional internal structure (e. g., monodromy, theta angles, or 2d gauge fields) can be realized by intersecting brane systems or M2-branes ending on M5-branes in M-theory. These constructions provide a direct link between field-theoretic data (such as the singularities of the gauge and scalar fields) and geometric data in the string dual, such as the embedding and fluxes on the probe branes.

%From the gravity side, the presence of surface operators breaks the full conformal symmetry of $AdS_5$  to that of $AdS_3$, corresponding to the preserved conformal symmetry along the defect. The resulting defect conformal field theory (dCFT) provides a controlled setting for studying correlation functions, operator insertions, and anomalies in the presence of extended objects. Holographic calculations allow one to extract observables such as one-point functions, entanglement entropy, and expectation values of the defect operator itself.

Our work will mainly focus on co-dimension 2 defects in the context of gauge/gravity dualities. 
The Gukov-Witten defect \cite{Gukov:2006jk} is probably the most known and simplest defect of co-dimension 2. 
It respects a $\mathfrak{psu}(1,1|2)^2\rtimes \mathfrak{su}(2)_R $ subalgebra of the full superconformal algebra $\mathfrak{psu}(2,2|4)$ and, as a result, it preserves half of the supersymmetries. In the dual  field theory, the defect is described by classical solutions having non-zero diagonal vevs for some of the scalar fields and have certain monodromies in the plane perpendicular to the defect. 
These vevs exhibit a singular behaviour on the defect, namely they have a simple pole (Nahm pole). The spacetime dependence of this singularity is dictated by conformal invariance. 
At strong coupling, the classification of $1/2$-BPS integrable boundary conditions of the string sigma model presented in \cite{Dekel:2011ja} contains this defect but only for the special case  in which the probe D3 brane has an inclination of $\pi/2$ with respect to the $AdS_5$
boundary.
Another $1/2$-BPS  co-dimension 2 defect is the one preserving a $\mathfrak{su}(1,1|4)\times \mathfrak{su}(1,1)$ subalgebra of the $\mathfrak{psu}(2,2|4)$ superalgebra. It is, most likely, realised by a D7 brane wrapping an $AdS_3\times S^5$ subspace \cite{Harvey:2008zz} of the $AdS_5\times S^5$ background. This defect completes the list of $1/2$-BPS defects of co-dimension 2. Among the non-supersymmetric defects which can be realised holographically, there is one that can be described by a flux stabilised 5-brane. It corresponds to a co-dimension 2 fuzzy $S^3$ in the classification of \cite{deLeeuw:2024qki}.  A defect that apparently does not belong in the classification of \cite{deLeeuw:2024qki} is the non-supersymmetric D3-D5 intersection presented in \cite{Georgiou:2025mgg}. Another recent study related to integrable non-supersymmetric defects based on the fuzzy $S^3$ is that of \cite{Gombor:2025qvk}.

This paper explores holographic realisations of non-supersymmetric co-dimension 2 surface operators in 
 ${\cal N}=4$ SYM, focusing on their construction via probe branes in type IIB string theory and their interpretation within the framework of AdS/dCFT correspondence. We comment on the worldvolume action, supersymmetry conditions, and physical observables associated with these defects.
 Our main interest will be on the holographic description of the duality. In this work, we will refrain from discussing  questions regarding  the integrability of the theories appearing on the two sides of the duality.\footnote{Recently, the integrability properties of the Gukov-Witten defects were studied in \cite{Holguin:2025bfe,Chalabi:2025nbg}. The conclusion is that although ordinary Gukov-Witten defects are not integrable except for special sub-sectors, the rigid Gukov-Witten defects are integrable, at the leading order, in a corner of their parameter space.} 
%This issue is left for future work.\footnote{Because there is nonzero flux present, we need to employ the generalized integrability framework introduced in \cite{LinardopoulosZarembo21, Linardopoulos22, Linardopoulos25a}.}

The plan of the paper is as follows. In section \ref{D5-sol}, we present the embedding of our D5 brane solution in the ambient $AdS_5\times S^5$ geometry. Our solution carries $k$-units of flux through an $S^2\subset S^5$. The symmetry of the induced metric on the D5 brane is $AdS_3\times S^1\times S^2$. The intersection of $AdS_3\times S^1$ with the boundary of $AdS_5$ creates the surface defect whose co-dimension is 2. The solution depends on two continuous parameters. One of them $\s$, determines the inclination angle of the D5 brane with respect to the boundary of the spacetime, while the other $\rho$ is related to the $S^1$ and determines the winding of the brane around one of the angles of the internal space $S^5$.
In section \ref{interpol}, we point out that classically our construction interpolates between the supersymmetric D3-D3 system at one end, and the non-supersymmetric D3-D5 system  presented in \cite{Georgiou:2025mgg} at the other.
In section \ref{stability}, we examine the stability of our configuration. In particular, we determine the regions of the parametric space of the parameters $(\rho,\s)$, that characterise our solution,
in which the masses of all the fluctuations of the transverse to the D5 brane coordinates respect the B-F bound. In section \ref{D5-D7}, we argue that for our construction to be valid one needs to insert in the geometry two D7 branes on which the D5 brane solution should terminate. This is necessary because the "winding" number $\rho$ is not an integer and as a result the D5 brane has boundaries. These boundaries induce a gauge anomaly for the D5 brane which cancels through anomaly inflow from the D7 branes.
The full system of the D5 and D7 branes is
anomaly free and has no tachyonic instabilities for a certain range of its parameters. 
In section \ref{dual}, we discuss the field theory dual CFTs of the D5-D7 system. We determine the classical solution of the ${\cal N}=4$ SYM equations of motion which we conjecture to describe the co-dimension 2 surface operators and comment on the quantities needed to describe these non-supersymmetric defects.
Finally, in section \ref{concl}, we present our conclusions and some possible future directions.

% the AdS3 has an arbitrary inclination angle with respect to the AdS5 boundary. The defect has a dual description as a probe D3 brane with geometry AdS3 × S1 ⊂ AdS5 × S5 [13, 18] where the intersection of AdS3 with the boundary of AdS5 generates the surface defect.

%%%%%%%%%%%%%%%%%%%%%%%%%%%%%%%%%%%%%%%%%%%%%%%%%%%%%%%%%%%%%%%%%%%%%%%%%%%%%%%%%%%%%

\section{The D5 probe brane}\label{D5-sol}

In this section, we focus on the gravity side and present the solution of a  novel D5-brane embedded in the $AdS_5\times S^5$ geometry. Our solution depends on two independent parameters and ends on a two-dimensional submanifold of the $AdS_5$ boundary.  It provides, thus, the realisation of the holographic dual of a co-dimension 2 defect conformal field theory (dCFT). 
The D5-brane wraps an $S^2$  of the internal space $S^5$ and extends along the $S^1\subset S^5$ parametrised by the angle $\tilde \gamma$. 
As a result, the symmetry of the induced, on the brane metric, is 
$AdS_3\times S^1\times S^2$.
The brane orientation of the co-dimension 2 D3-D5 probe-brane system 
%(see \eqref{metric} for the metric and \eqref{embedding} for the embedding ansatz of the D5-brane) 
is given in table \ref{Table:C2D3D5system}, with the worldvolume coordinates of the D5-brane being $\zeta^\mu = (x_0,x_1,r,\tilde \gamma, \beta, \gamma)$. 
\begin{table}[H]
\begin{center}\begin{tabular}{|c||c|c|c|c|c|c|c|c|c|c|}
\hline
& ${\color{red}x_0}$ & ${\color{red}x_1}$ & ${\color{red}r}$ & ${\color{red}\psi}$ & ${\color{red}z}$ & ${\color{blue}\tilde \psi}$ & ${\color{blue} {\tilde \beta} }$ & ${\color{blue} {\tilde{ \gamma}}}$ & ${\color{blue}\beta}$ & ${\color{blue}\gamma}$ \\ \hline
\text{D3} & $\bullet$ & $\bullet$ & $\bullet$ & $\bullet$ &&&&&& \\ \hline
\text{D5 probe} & $\bullet$ & $\bullet$ & $\bullet$ &  &   & & &$\bullet$ & $\bullet$ & $\bullet$ \\ \hline
\text{D7 probe} & $\bullet$ & $\bullet$ & $\bullet$ & &$\bullet$   & &$\bullet$&$\bullet$ & $\bullet$ & $\bullet$ \\ \hline
\end{tabular}
\caption{The D3-D5-D7 intersection. The D7 brane and its role are described and analysed in section \ref{D5-D7}.\label{Table:C2D3D5system}}\end{center}
\end{table}

%%%%%%%%%%%%%%%%%%%%%%%%%%%%%%%%%%%%%%%%%%%%%%%%%%%%%%%%%%%%%%%

\subsection{Embedding ansatz of the D5 probe brane} 
\label{emb_ansatz}

In what follows, we will present our solution for the D5-brane which realises the gravity duals of certain non-supersymmetric 
co-dimension 2 defect conformal field theories in the case of Lorentzian signature.

Our starting point will be the action
\begin{equation}
\label{D5-Lor}
S_{D5}=- \frac{T_5}{g_s}\Bigg\{\int d^6 \zeta \sqrt{-{\rm det}\, \mathcal P [g+2 \pi \a' F]}-
2 \pi \a' \int \mathcal P [ F\wedge C_4]\Bigg\} 
\end{equation}
where  $\mathcal P$ denotes the pullback of the different spacetime fields on the worldvolume of the brane, $T_5$ is the tension of the D5-brane and $F$ is the field strength of the worldvolume gauge field. The tension of the brane  and the string coupling are given by
\begin{equation} \label{def-tension-coupling}
T_5 = \frac{\lambda^{3/2}}{(2 \, \pi)^5} \quad \& \quad g_s = \frac{g_{YM}^2}{4\, \pi} \quad {\rm with} \quad \lambda = \alpha'^{-2} \, . 
\end{equation}
Notice that we work in units in which the radius of the $AdS_5$ 
is taken to be 1. In \eqref{D5-Lor} we have also used the fact  that the 
$B$-field vanishes for the $AdS_5\times S^5$ solution of the 
type IIB supergravity equations.

The metric of the background written in Poincare coordinates is given by
\begin{equation}\label{metric-Lor}
ds^2 = \frac{1}{z^2} \, \Big[-dx_0^2 + dx_1^2+ dr^2 + r^2 \, d\psi^2 + dz^2 \Big] +d\Omega_5^2
\end{equation}
with the length element on $S^5$ being
\begin{equation}\label{metric}
d\Omega_5^2 = 
d\tilde\psi^2 + 
\sin^2 \tilde\psi \, \left(d{\tilde \beta}^2 + \sin^2 {\tilde \beta} \, d{\tilde \gamma}^2 \right) +
\cos^2 \tilde\psi \left(d\beta^2 + \sin^2 \beta \, d\gamma^2\right) \, . 
\end{equation}
In addition, we choose the RR 4-form potential to be given by
\begin{equation}\label{C4}
C_4 = - \frac{r}{z^4} \, dx_0 \wedge dx_1 \wedge dr \wedge d\psi \, . 
\end{equation}
As mentioned above, we have chosen the world-volume coordinates of a D5-brane to be $x_0,x_1,r,\tilde \gamma,$ $ \beta, \gamma$ and consider the following ansatz for the rest of the coordinates
\begin{equation}\label{embedding}
\tilde\psi = \tilde\psi_0 \, , \quad 
{\tilde \beta} = \frac{\pi}{2} ,\quad 
{\psi} =\rho \,\tilde \gamma +\phi_0\quad \& \quad
z= \sigma \, r \, . 
\end{equation}
%
%In figure \ref{figg-1} we depict the D5 brane with the coordinates $x_0, x_1, \beta$ and $\gamma$ suppressed. 
%Let us make a brief comment on  the relation between
%the parameter  $\s$ of our solution and the physical parameters $k$ and $\lambda$. In the limit of small $\s$  the flux $k$ becomes very large and the inclination of the brane with respect to the $AdS_5$ boundary goes to zero. On the other hand, when $\s=1$, which is the endpoint of the range of validity of our solution, the inclination angle becomes $\pi/4$ (see figure   \ref{figg-1}).

The solution is supported by a worldvolume 2-form flux through the $S^2$ parametrised by the angles $(\beta,\gamma)$. The corresponding  1-form gauge field living on the world-volume of the D5-brane is given by
\begin{equation}\label{A}
A = \frac{\kappa}{2 \, \pi \, \alpha'} \, \cos \beta \, d\gamma .
\end{equation} 
Its functional form is determined by the requirement that the equation of motion for the potential $A$ is satisfied.
One can check that, given the ansatz in \eqref{embedding}, the equations of motion derived from the action \eqref{D5-Lor} 
are satisfied automatically for the coordinates $\psi$ and $\tilde \beta$. The equation of motion of the coordinate $\tilde \psi$ determines the value of the constant $\kappa$ to be
\begin{equation}\label{kappa}
\kappa = \frac{1}{\sigma} \, \cos\tilde \psi_0 
\, \sqrt{\big. \left(2-3 \cos^2\tilde \psi_0 \right) 
\s^2 +2 \rho ^2}\, .
\end{equation}
Finally, the equation of motion for the holographic coordinate $z$ forces the constant angle $\tilde \psi_0$ to take the following value
\begin{equation}\label{tildepsi}
\cos^2\tilde \psi_0= \frac{2 \, \sqrt{2}}{ 3 \, \sigma} 
\, \rho \, \sqrt{ \big. \left(\rho ^2-1\right) 
\left(1+\sigma ^2\right)}- (\rho ^2-1)\, .
\end{equation} 
Consequently, we see that our D5-brane is characterised by two independent parameters, namely $\sigma$ and $\rho$. 
The first determines the inclination of the brane with respect to the 
boundary of the $AdS_5$ space and it was also present in the construction of \cite{Georgiou:2025mgg}, 
as well as in the supersymmetric D3-D3 system \cite{Drukker:2008wr}.  
The second parameter, $\rho$, is the new ingredient of our solution and, as can be seen from \eqref{tildepsi}, is crucial in determining  the size of the $S^2$ around which the brane wraps.

\begin{figure}[h!]
 \centering
  \includegraphics[width=8cm,height=8cm]{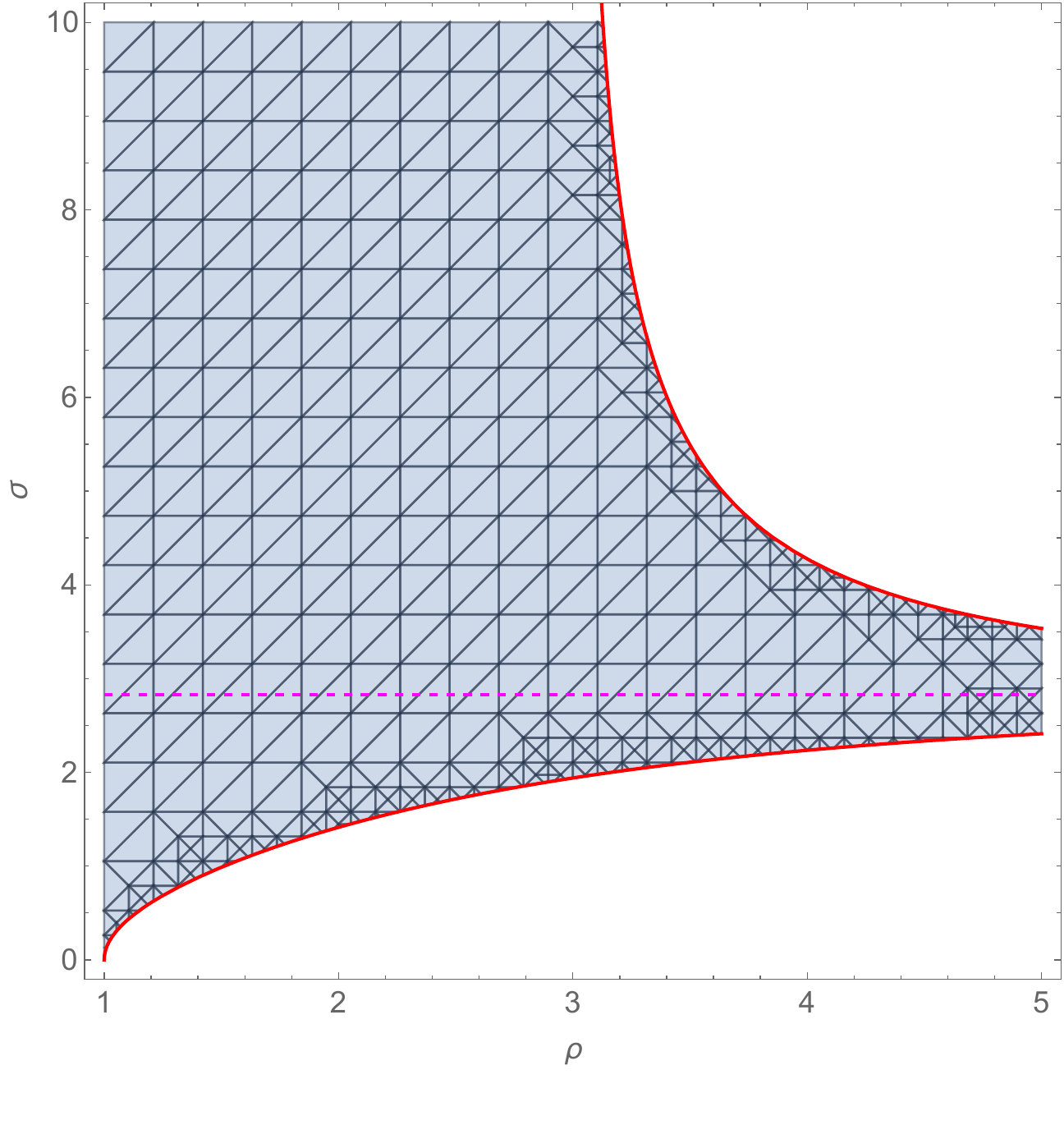}
 \caption{A graph depicting the portion of the  parametric space 
 $(\sigma,\rho)$ for which our solution is valid. 
 The two red lines correspond to the limiting curves that 
 are presented in equation \eqref{boundss} 
 and the dotted magenta line is for the value $\sigma = 2\sqrt{2}$.}
 \label{figg-1}
 \end{figure}
 
Let us now comment on the range in which the parameters $\rho$ and $\s$ 
can take values. Without any loss of generality, we will assume that 
$\rho\geq 0$. The condition $0\leq \cos{\psi_0}\leq 1$ implies the following relations between  $\rho$ and $\s$
 \begin{eqnarray}\label{boundss}
&&1 \leq \rho \leq \sqrt{\frac{8 \sigma ^2+8}{8-\sigma ^2}}\qquad {\rm if} \qquad 0<\sigma <2 \, \sqrt{2} \nonumber \\[5pt]
&& 1\leq \rho \leq 3 \, \sqrt{\frac{\sigma ^2}{\sigma ^2-8}} \qquad {\rm if} \qquad \sigma >2 \, \sqrt{2}.
\end{eqnarray}
Thus, we see that the value of the parameter $\rho$ is restricted to be lower than a certain value both for $\s<2 \, \sqrt{2}$ and $\s>2 \, \sqrt{2}$.
These values are depicted in  figure \ref{figg-1}. The shaded area depicts the range in which our solution is valid. 

 \begin{figure}[h!]
 \centering
  \includegraphics[width=9cm,height=9cm]{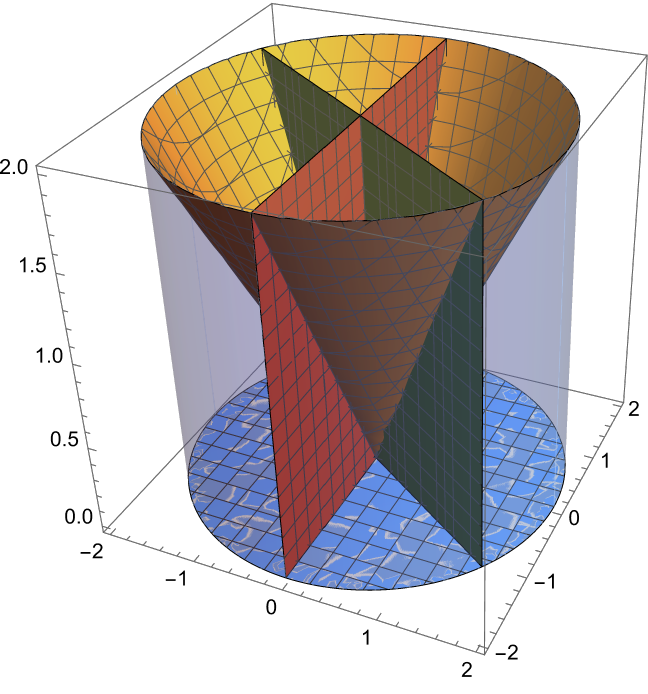}
 \caption{A picture depicting the D5 brane. The yellow cone is the D5 brane.
 The point at which the cone touches the boundary $z=0$ is really a two dimensional surface since the coordinates $x_0$ and $x_1$, along which the brane extends, have been suppressed.
 The blue horizontal plane is parametrised by the coordinates $(x_2,x_3)=(r \cos{\psi},r \sin{\psi})$ and represents the boundary of the $AdS_5$ space with the vertical axis being the holographic coordinate $z$. The green and red vertical planes represent the two D7 branes. The D5 brane has one of its boundaries on the green D7 brane and the other on the red one. The D5 starts from the red D7, winds anticlockwise once or more times around the $\psi$ angle and ends on the green D7 brane. The necessity of the D7 branes is discussed in section \ref{D5-D7}. }
 \label{brane-cone}
 \end{figure}
 
At this point, let us comment on the co-dimension of the defect that our D5-brane introduces. Notice that the D5-brane intersects the $AdS_5$ boundary at $z=r=0$, implying that the defect is a two-dimensional plane $\mathbb R^{(1,1)}$ that extends along two of the four directions of the boundary, that is, along $x_0$ and $x_1$. As a result, we have a defect of co-dimension 2.
To quantise the flux we impose the following condition
\begin{equation}\label{kappa-1}
k=-\int_{S^2} \frac{F}{2 \, \pi} \quad \Rightarrow \quad 
k=\frac{1}{2 \pi}\frac{ \kappa}{2 \pi \alpha'}\int_0^\pi\sin \beta\,d\beta \int_0^{2\pi}d\gamma \quad \Rightarrow \quad k=\frac{\kappa}{\pi \a'}=\frac{\kappa \sqrt{\lambda}}{\pi}\, ,
\end{equation}
with $k \in {\mathbb N}^*$.
It should be emphasized that
our D5-brane solution breaks all supersymmetries, i.e. there is no Killing 
spinor of the $AdS^5\times S^5$ background that is preserved in the presence of 
the probe brane. This fact is in agreement with the dual field theory picture. The corresponding field theory analysis is presented in appendix \ref{Appendix:Supersymmetry}.

Before closing this section it is instructive to write down the metric induced on the D5 brane world-volume. This reads
\be\label{induced-metric}
ds^2_{ind}=\frac{1}{r ^2 \s^2}
\Bigg[-dx_0^2+dx_1^2+\left(1+\s^2\right)dr^2\Bigg]+ 
\left[\sin^2\tilde \psi_0+\frac{\rho^2}{\s^2}\right] 
d\tilde\gamma^2+\cos^2\tilde \psi_0 
\left(d\beta^2+ \sin^2\beta\, d\gamma^2 \right) \, .
\ee
After performing the change of variables $r=\frac{\hat r}{\sqrt{1+\s^2}}$ the metric above becomes
\be\label{induced-metric-1}
ds^2_{ind}=\frac{1+\s^2}{\s^2}\,
\frac{-dx_0^2+dx_1^2+d\hat r^2}{\hat r^2}+
\left[\sin^2\tilde \psi_0+\frac{\rho^2}{\s^2}\right]d\tilde\gamma^2 
+\cos^2\tilde \psi_0 \left(d\beta^2+ \sin^2\beta d\gamma^2 \right).
\ee
In this last equation, the $AdS_3\times S^1\times S^2$ symmetry of the D5 brane becomes apparent. Finally, for completeness we write down the on-shell value of the D5 brane Lagrangian density which is given by
\be\label{Laf-on-shell}
{\cal L}_{on-shell}=-\frac{1}{r^3 \, \sigma ^5} \, 
\sin \beta  \Bigg[\sqrt{2}\cos\tilde \psi_0 \, \sqrt{\big. 1+\sigma ^2} \,  
\left(\rho^2+\sin^2\tilde \psi_0 \,\sigma ^2\right)-
\kappa\, \rho\, \s \Bigg] \, . 
\ee
The first term in the parenthesis is the DBI contribution while the second term is the WZ part.
%%%%%%%%%%%%%%%%%%%%%%%%%%%%%%%%%%%%%%%%%%%%%%%%%%%%%%%%%%%%%%%%%%%%%%%%%%%%

\subsection{Holographic interpolations}\label{interpol}
%In this section we comment on the dependence of our solution on the parameter $\l$.
As mentioned above, our brane solution depends on two independent parameters $\s$ and $\rho$. As can be seen from \eqref{tildepsi} a particular combination of these determines the value of the angle $\tilde \psi_0$. In turn, this value determines the radius of the $S^2$, which is $\cos{\tilde \psi_0}$, as well as the radius of the $S^1$ which is $\sin{\tilde \psi_0}$. We, thus, see that as the radius of $S^2$ increases the radius of the $S^1$ decreases, and vice versa. There are two endpoints. The first endpoint is when $\cos{\tilde \psi_0}=1$ in which case the radius of the $S^2$ becomes maximal while the $S^1$ shrinks to a point. This happens when the upper bound in \eqref{boundss} is saturated, namely when $\rho=\sqrt{\frac{8 \sigma ^2+8}{8-\sigma ^2}}$, for example. By taking into account that $\psi=\rho \,\tilde \gamma+ \phi_0$, it is easy to find that the induced on the brane metric in that limit becomes 
%where $\rho \rightarrow \sqrt{\frac{8 \sigma ^2+8}{8-\sigma ^2}}$. 
%From \eqref{induced-metric} one gets 
\be\label{induced-metric-limit1}
ds^2_{ind}=\frac{1}{r ^2 \s^2} 
\Bigg[-dx_0^2+dx_1^2+(1+\s^2)dr^2 \Bigg] 
+\frac{1}{\s^2} \,  d\psi^2+\left(d\beta^2+ \sin^2\beta\, d\gamma^2\right) \, .
\ee
But this is the induced metric for the brane solution of \cite{Georgiou:2025mgg} (see equation (2.10)). Furthermore, as can be seen from \eqref{kappa}, the strength of the gauge field $A$ in that limit becomes 
\be
\kappa=\frac{4+\s^2}{\s \sqrt{8-\s^2}} \, , 
\ee
which is precisely the value of the strength of the gauge field in the construction of \cite{Georgiou:2025mgg} (see equation (2.7)).
We conclude that in the aforementioned limit our D5-brane solution becomes that of \cite{Georgiou:2025mgg}.

The second endpoint is a bit more intricate. It can be reached by sending $\rho \rightarrow 1$. In this case, it can be seen from \eqref{tildepsi}  that $\cos{\tilde \psi}_0\rightarrow 0$. 
Consequently, it is now the  size of the two-sphere parametrised by $(\beta,\gamma)$ that goes to zero. In this limit two of the directions of the brane shrink to zero and the D5 brane loses two of its coordinates. The D5 solution resembles then that of a D3-brane. This can be easily seen from \eqref{induced-metric}  which, in the limit $\rho=1$, becomes
\be\label{induced-metric-limit2}
ds^2_{ind}=\frac{1}{r ^2 \s^2}
\Bigg[-dx_0^2+dx_1^2+(1+\s^2)dr^2\Bigg]+
\left(1+\frac{1}{\s^2}\right) d\tilde\gamma^2,
\ee
while the solution itself becomes
\begin{equation}\label{embedding-D3}
\tilde\psi = \frac{\pi}{2} \, , \quad 
{\tilde \beta} = \frac{\pi}{2} ,\quad 
{\psi} =\tilde \gamma +\phi_0\quad \& \quad
z= \sigma \, r \, . 
\end{equation}
Furthermore, in this limit one gets $\kappa\rightarrow 0$ which is consistent with the fact that the size of the $S^2$ goes to zero. From \eqref{induced-metric-limit2} one can see that the induced metric on the D5 brane becomes degenerate and the DBI contribution to the action vanishes.
Now eq. \eqref{embedding-D3} is precisely the solution of the supersymmetric D3 probe brane which realises the gravity dual of the Gukov-Witten surface operators \cite{Drukker:2008wr,Gukov:2006jk}.

At this point let us stress that, in the limit $\tilde\psi_0\rightarrow \frac{\pi}{2}$ which we consider here,  the object in our classical probe analysis is still a D5-brane, but in a collapsed, singular configuration whose classical worldvolume geometry  and supersymmetry coincide with those of the 1/2-BPS D3-brane. It is not literally a D3-brane.  The two objects will be distinguish when quantum ($\a'$ or $g_s$) 
corrections will be taken into account. However, since we are working at the classical limit we will  abuse language and refer to that solution as the 1/2-BPS D3 brane that is dual to the Gukov-Witten surface operators. It would be interesting to understand if in the $\tilde\psi_0\rightarrow \frac{\pi}{2}$ limit there exists  a mechanism which may dynamically turn the D5 brane into a  real D3 brane.

To recapitulate, our construction provides the gravity duals of a class of solutions that interpolates between the supersymmetric Gukov-Witten surface operators and the non-supersymmetric defect CFTs 
presented in \cite{Georgiou:2025mgg}.

Our D5 probe brane breaks all supersymmetries of the ambient spacetime $AdS_5\times S^5$. However, in the limit $\rho\rightarrow 1 \Rightarrow \cos\tilde\psi_0=\tilde \epsilon\rightarrow 0$ supersymmetry gets restored and one obtains a 1/2 BPS object in agreement with the fact that the worldvolume geometry of the D5 brane becomes that of the supersymmetric D3 brane. 
It will be enough to show that the projector of the D5 brane becomes identical to that of the 1/2 BPS D3 brane, the latter can be found in  \cite{Drukker:2008wr}. The situation is a bit tricky because, as mentioned above, the DBI Lagrangian becomes zero on-shell. The equation that determines the preserved Killing  spinors $\epsilon$ is 
\be\label{Killing}
 \Gamma_{\kappa}\epsilon |_{D5}=\epsilon |_{D5},
\ee
where in the case of our D5 brane the projection matrix is given by \cite{Skenderis:2002vf}
\begin{equation}\label{proj}
 \Gamma_{\kappa} =\frac{i}{\cL^{}_{DBI}} \, 
 \partial_0x^\mu\,\partial_1 x^\nu\,\partial_\omega x^\rho\,
\partial_{\tilde\gamma} x^\sigma 
 \Bigg[-
\partial_\beta x^\tau\partial_\gamma x^\l\,\Gamma_{\mu\nu\rho\sigma\tau \l} \, K+\,
\hat F \,\Gamma_{\mu\nu\rho\sigma}\Bigg]\, ,
\end{equation}
where $K$ acts on spinors as $K\psi=\psi^\star$ and $\hat F=-\kappa \sin\beta$ the strength of the worldvolume gauge field that lives on the D5 brane.

%%%%%%%%%%%%%%%%%%%%%%%%%%%%%%%%%%%%%%%%%%%%%%%%%%%%%%%%%%%%%%%%%%%%%%%%%%%%%%%%%%%%%%%%%

As mentioned above at $\tilde\psi_0=\frac{\pi}{2}$, $\cL_{DBI}=0$.
To make sense of equation \eqref{Killing} one should expand both the DBI Lagrangian density and the gauge field $F$ in powers of $\tilde \epsilon=\cos\tilde\psi_0$ and keep the finite term, hoping that there is no divergent term.
Before doing so let us comment on equation \eqref{proj}. In order to make a direct comparison with the projection matrix of \cite{Drukker:2008wr} we should change the embedding coordinates of our D5 brane solution from $\zeta^\mu = (x_0,x_1,r,\tilde \gamma, \beta, \gamma)$ to $\zeta^\mu = (x_0,x_1,\omega,\tilde \gamma, \beta, \gamma)$, where $\omega=\frac{1}{z}$. Then the equation $z=\s r$ translates to $r=\frac{1}{\omega \s}=\frac{\sinh u_0}{\omega} $ which is the relation appearing in  \cite{Drukker:2008wr}. After this change of variables the gauge field and Lagrangian density of our D5 solution become
\begin{equation}
\hat F= -\tilde \epsilon \, \frac{\sqrt{2(1+\s^2)}}{\s} \sin\beta+\mathcal{O}(\tilde\epsilon^2)
\quad {\rm and} \quad  
\cL_{DBI}= \tilde\epsilon\,\omega \, \frac{\sqrt{2(1+\s^2)^3}}{\s^3} \sin\beta + \mathcal{O}(\tilde\epsilon^2) \, .
\end{equation}
The first term on the right hand side of \eqref{proj} goes to zero in the limit $\tilde\epsilon\rightarrow 0$. This is so because  two of the flat gamma matrices come multiplied with the vielbeins of the coordinates $\beta$ and $\gamma$, that is
multiplied by $\cos\tilde\psi_0$ each. As a result the numerator of this first term is proportional to $\tilde\epsilon^2$ while the denominator is proportional to $\tilde\epsilon$ resulting to a vanishing contribution of the first term in the limit $\tilde\epsilon \rightarrow 0$.
The second term in \eqref{proj} gives
\begin{equation}\label{proj-1}
 \Gamma_{\kappa}
%= \frac{-i \, \s^2}{\omega (1+\s^2)}\,\partial_0 x^\mu\,\partial_1 x^\nu\,\partial_\omega x^\rho\,
%\partial_{\tilde\gamma} x^\sigma\,\Gamma_{\mu\nu\rho\sigma}
=
\frac{- \, i}{\omega \cosh^2u_0}\,\partial_0 x^\mu\,\partial_1 x^\nu\,\partial_\omega x^\rho\,
\partial_{\tilde\gamma} x^\sigma\,\Gamma_{\mu\nu\rho\sigma}\, ,
\end{equation}
which is identical to equation (C.9) of \cite{Drukker:2008wr}. In \eqref{proj-1} $\omega \cosh^2u_0$ is the DBI Lagrangian density of the D3 brane.
We conclude that the preserved supersymmetries are the same as those preserved by a 
supersymmetric  surface operator in $\cN=4$ SYM which preserves 16 supersymmetries.
The corresponding projectors are \cite{Drukker:2008wr}
\begin{equation}
\left(1+\gamma_{2389}\right)\epsilon_1 \, = \, 0 \quad {\rm and} \quad 
\left(1+\gamma_{2389}\right)\epsilon_2 \, = \, 0 \, . 
\end{equation}
Needless to say that the $\kappa$-symmetry analysis above agrees with the corresponding field theory analysis. 
%%%%%%%%%%%%%%%%%%%%%%%%%%%%%%%%%%%%%%%%%%%%%%%%%%%%%%%%%%%%%%%%%%%%%%%%%%%%%%

\section{B-F bound and absence of tachyonic instabilities}\label{stability}

%\begin{figure}[h!]
 %\centering
  %\includegraphics[width=0.9\textwidth]{ms2-sm.pdf}
 %\caption{A graph depicting the upper bound which the parameter $\l$ can take as a function of the inclination $\s$. The  $\l$  axis is the vertical one while the $\s$ axis is the horizontal. The portion of the  parametric space $(\l,\s)$ for which our solution is valid is the area below the curves.}
 %\label{figg-1}
 %\end{figure}

%\begin{figure}[h!]
% \centering
%  \includegraphics[width=0.9\textwidth]{ms2-big.pdf}
% \caption{A graph depicting the upper bound which the parameter $\l$ can take as a function of the inclination $\s$. The  $\l$  axis is the vertical one while the $\s$ axis is the horizontal. The portion of the  parametric space $(\l,\s)$ for which our solution is valid is the area below the curves.}
% \label{figg-1}
% \end{figure}
In this section, we will examine the stability of the defect D5-brane presented in section \ref{emb_ansatz}.
As mentioned in the previous section, the induced metric on the D5-brane worldvolume is $AdS_3\times S^1\times S^2$. The $S^1$ corresponds to the isometry of the $S^5$, $\tilde \gamma$, while the $S^2$ corresponds to the 2-sphere around which the D5 probe brane wraps. The stability of the solution is ensured if the fluctuations of the coordinates transverse to the brane acquire masses which saturate or are above the Breitenlohner-Freedman (B-F) bound, namely $m^2\ge m^2_{BF}$ \cite{Breitenlohner:1982bm}. From the point of view of the D5-brane the transverse fluctuations behave as scalars propagating on the hyperbolic space with geometry $AdS_3$. For a space-time with geometry $AdS_{d+1}$ of radius $\ell$, the B-F bound turns out to be $m^2_{BF}=-\frac{d^2}{4 \ell}$. 
We are using units in which  $\ell=1$. As a result, our solution will have no tachyonic instabilities  if the masses of the fluctuations satisfy $m^2\ge -1$. In what follows, we will see that the B-F bound will impose  further constraints on the parametric space $(\l,\s)$ besides those imposed by the equations of motion (equation \eqref{boundss}).

We now introduce fluctuations around the D5 brane solution both for the world-volume gauge field 
$A= \left\{\kappa \cos \beta + \d A(\zeta^{\mu})\right\} d\gamma $, as well as for transverse coordinates, i. e.
\begin{equation}
z=\s \, r\big(1+ \d z(\zeta^{\mu})\big)\, , \quad 
{\tilde \psi} = \tilde \psi_0 + \d {\tilde \psi} (\zeta^{\mu})\, , \quad 
{\tilde \beta} = \frac{\pi}{2}+\d {\tilde \beta}(\zeta^{\mu}) \, , \quad 
{\psi}= \rho \,  \tilde \gamma+\phi_0+\d {\psi} (\zeta^{\mu}) \, . 
\end{equation}
The next step is to plug these relations into the action \eqref{D5-Lor} and keep terms which are at most quadratic in the fluctuations.
As usual, the terms linear in the fluctuations vanish, since we are expanding around a solution of the equations of motion. 
The resulting Lagrangian is then used in order to
derive the equations governing the time evolution of the fluctuations. From them one can determine the mass of each fluctuation and check  under which conditions these masses satisfy the B-F bound.

%Bearing in mind that  we are primarily interested in the time evolution of the %fluctuations, we  make the simplifying assumption that the fluctuations depend only %on the time coordinate $x_0$.
The equation of motion for the fluctuations around $\psi= \rho\,  \tilde \gamma+\phi_0$, namely  $\d { \psi}(x_0)$ gives
\begin{equation} \label{dpsi}
 \d { \psi}''(x_0)=0 \Leftrightarrow  \d { \psi}''(s)=0\Rightarrow m^2_\psi=0>-1\equiv m^2_{BF},
\end{equation}
where as usual we passed from the time $x_0$ to the proper time $s$, by the use of  $ds=\frac{dx_0}{z}=\frac{dx_0}{\s \,r}$.
Similarly, the equation for the fluctuations of the gauge field give
\begin{equation} \label{dA}
\d A''(x_0)=0\Rightarrow m^2_A=0>-1\equiv m^2_{BF}\, .
\end{equation}
We now turn to the equation for the fluctuations of $\tilde \beta$. This reads
\begin{equation} \label{dtilde-beta}
\left( \sin ^2\tilde \psi_0+\frac{\rho ^2}{\s^2}\right)  \d {\tilde \beta}''(s)-\d {\tilde \beta}(s)=0\Rightarrow m^2_{\tilde \beta}=-\frac{\sigma ^2}{\sigma ^2 \sin ^2\tilde \psi_0+\rho ^2}\, .
\end{equation}
By demanding $m^2_{\tilde \beta}\geq -1$, we deduce that 
for $\rho \geq 3$ there is no restriction on $\rho$ and $\s$ while for 
$1 < \rho \le 3$ the following restriction emerges 
\begin{equation} \label{co-2}
\rho  \sqrt{\frac{1}{9-\rho ^2} \, 
\Bigg[2 \sqrt{2} \sqrt{\frac{2 \, \rho ^2+7}{\rho ^2-1}} +5 \Bigg]}>\sigma \, .
\end{equation}
To proceed notice that in the range $0<\sigma \le 2\sqrt{2}$ the  first condition in \eqref{boundss} can be inverted to give
\be\label{inversebound}
\frac{2 \sqrt{2} \sqrt{\rho ^2-1}}{\sqrt{\rho ^2+8}} \le \sigma \leq 
2\sqrt{2} \, .
\ee
Thus we see that, as long as 
$\rho  \sqrt{\frac{1}{9-\rho ^2} \, 
\Bigg[2 \sqrt{2} \sqrt{\frac{2 \, \rho ^2+7}{\rho ^2-1}} +5 \Bigg]}\geq2\sqrt{2}$,
then equation \eqref{co-2} puts no constraint on the domain of validity of 
our solution. This happens when $\rho$ is in the range
\begin{equation}
\rho \leq2 \sqrt{\frac{2}{17} \left(5-2 \sqrt{2}\right)}=1.0109 
\quad {\rm and} \quad
\rho \geq 2 \sqrt{\frac{2}{17} \left(2 \sqrt{2}+5\right)}=1.9194 \, 
\end{equation}
When $\rho \in(1.0109,1.9194)$ instead of \eqref{inversebound} we have the more severe constraint
\be\label{inversebound-severe}
\frac{2 \sqrt{2} \sqrt{\rho ^2-1}}{\sqrt{\rho ^2+8}} \le \sigma \le 
\rho  \sqrt{\frac{1}{9-\rho ^2} \, 
\Bigg[2 \sqrt{2} \sqrt{\frac{2 \, \rho ^2+7}{\rho ^2-1}} +5 \Bigg]}
\quad {\rm for} \quad \rho \in(1.0109,1.9194).
\ee
To recap, we have seen that the requirement that the mass for the fluctuations of $\tilde \beta$ obeys the B-F bound imposes a mild  constraint \eqref{inversebound-severe} besides those already imposed by the equations of motion  that are written explicitly in \eqref{inversebound}.

\begin{figure}[h!]
 %\centering
  %\includegraphics[width=.62\textwidth]{fifi1-1.pdf}
  %\includegraphics[width=0.52\textwidth]{ms2-smaller.pdf}
  \includegraphics[width=7.5cm,height=7cm]{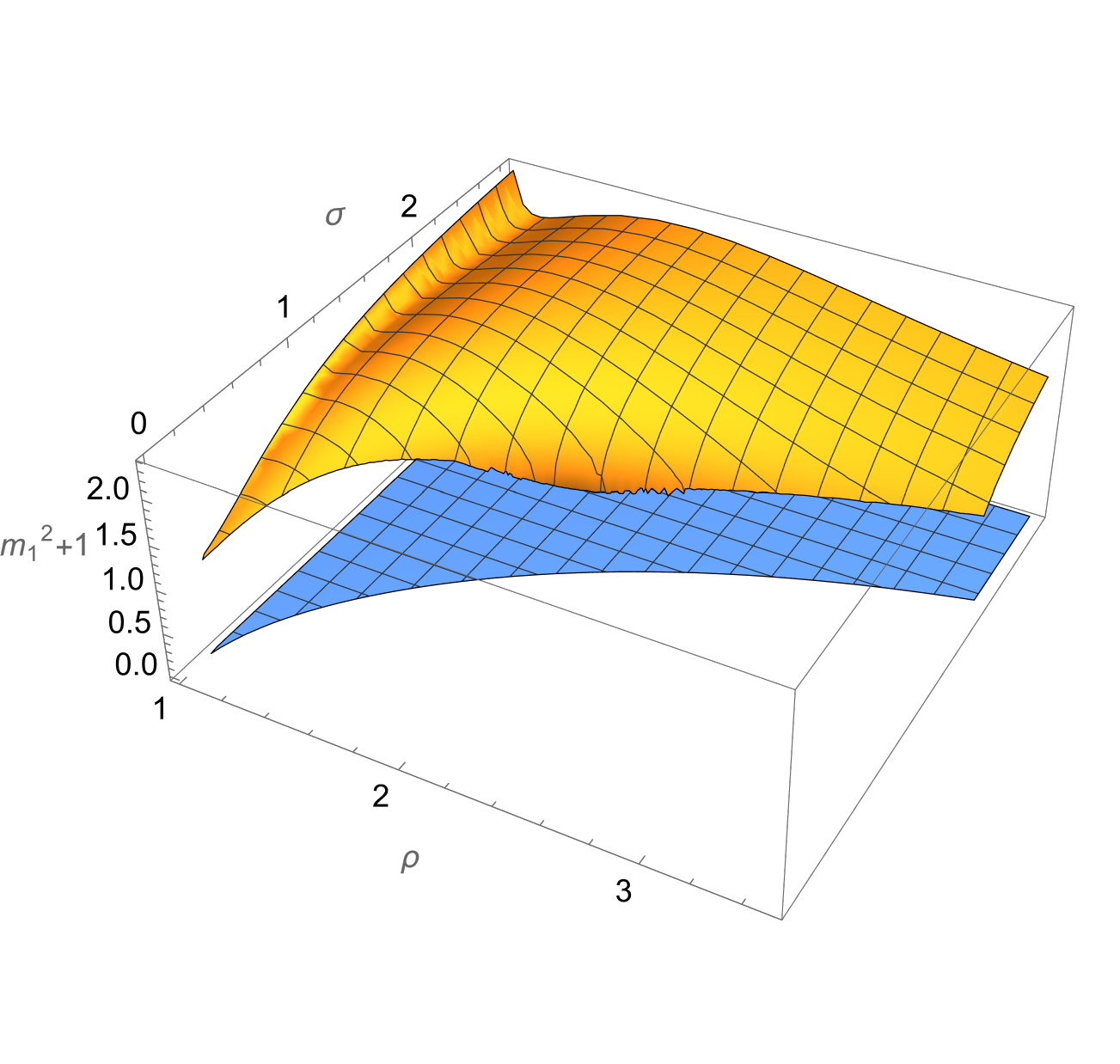}
    \includegraphics[width=7.5cm,height=7cm]{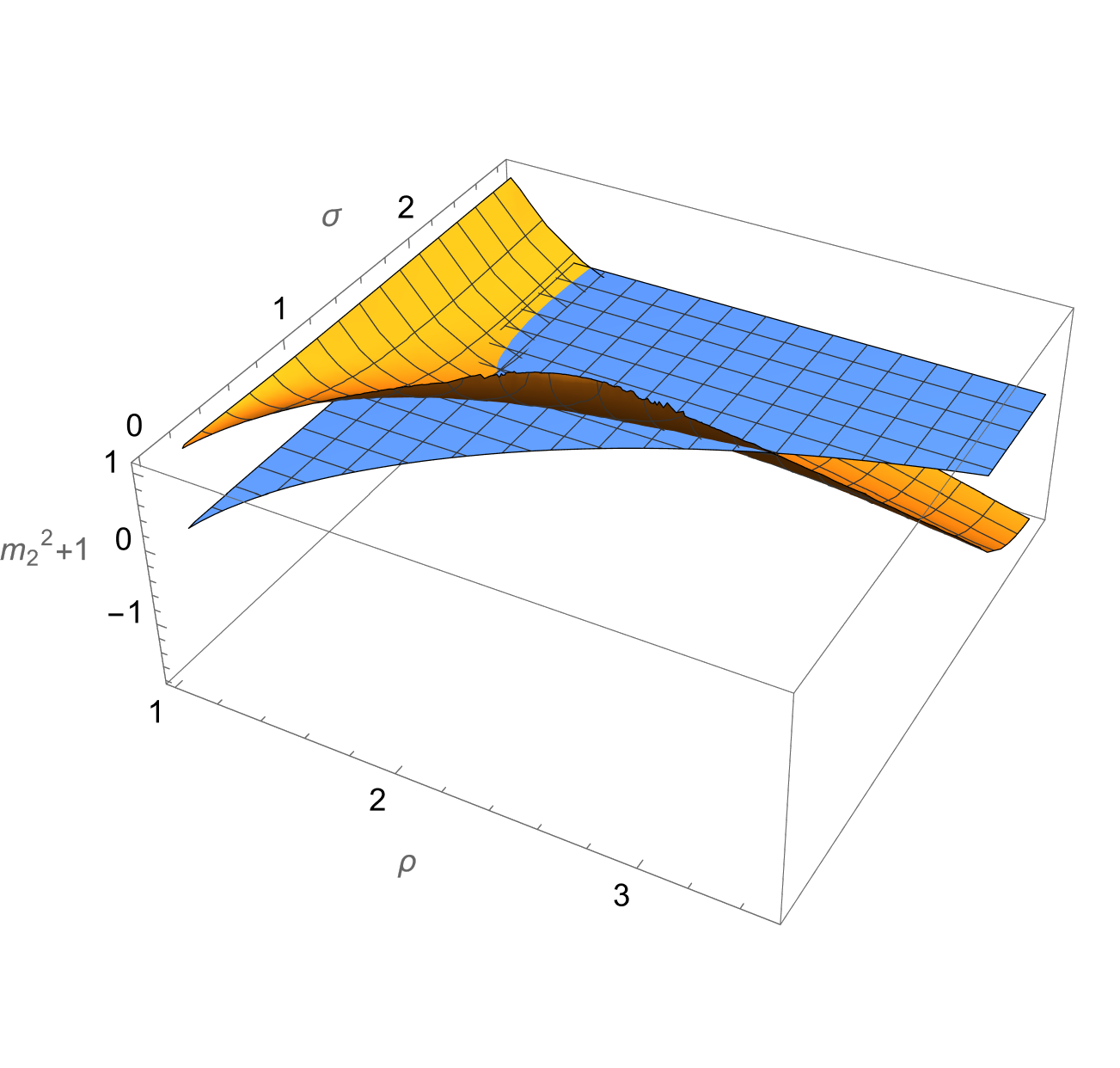}
 \caption{On the left graph we have drawn the function $m^2_1(\rho,\s)+1$ for the range of $\s \in [0,2 \sqrt{2}]$ subject to the condition \eqref{boundss}. We have allowed $\rho$ to be between $1$ and $3.5$. The blue plane is the zero of the vertical axis $z=0$. On the right graph we have drawn the function $m^2_2(\rho,\s)+1$ for the same values of $\rho$ and $\s$.}
 \label{figg-2}
 \end{figure}
 
 \begin{figure}[ht]
 %\centering
  %\includegraphics[width=0.45\textwidth]{ms2-smaller-region.pdf}
   %\includegraphics[width=0.45\textwidth]{ms2-smaller-region-restricted.pdf}
   \includegraphics[width=7.5cm,height=7cm]{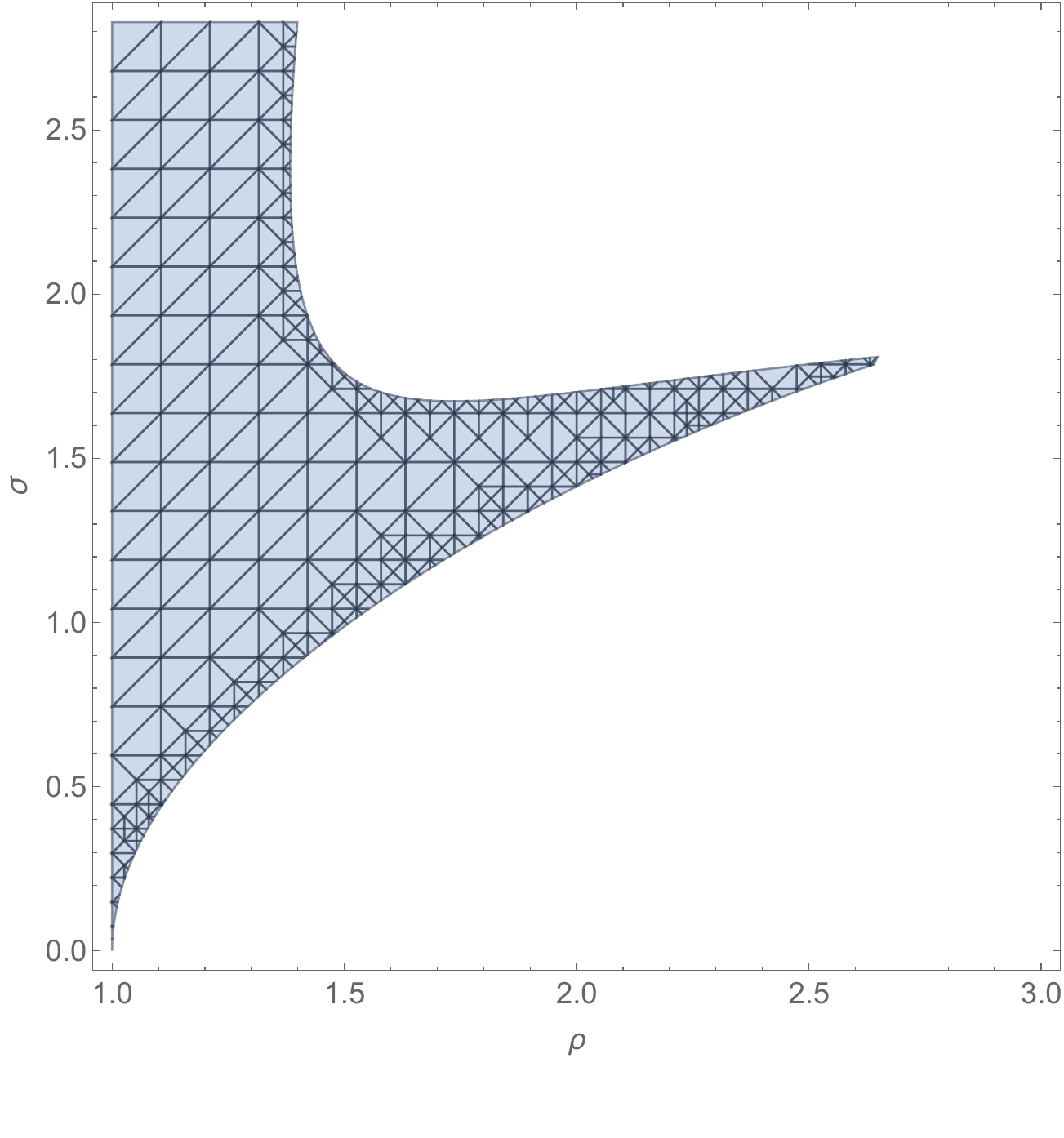}
    \includegraphics[width=7.5cm,height=7cm]{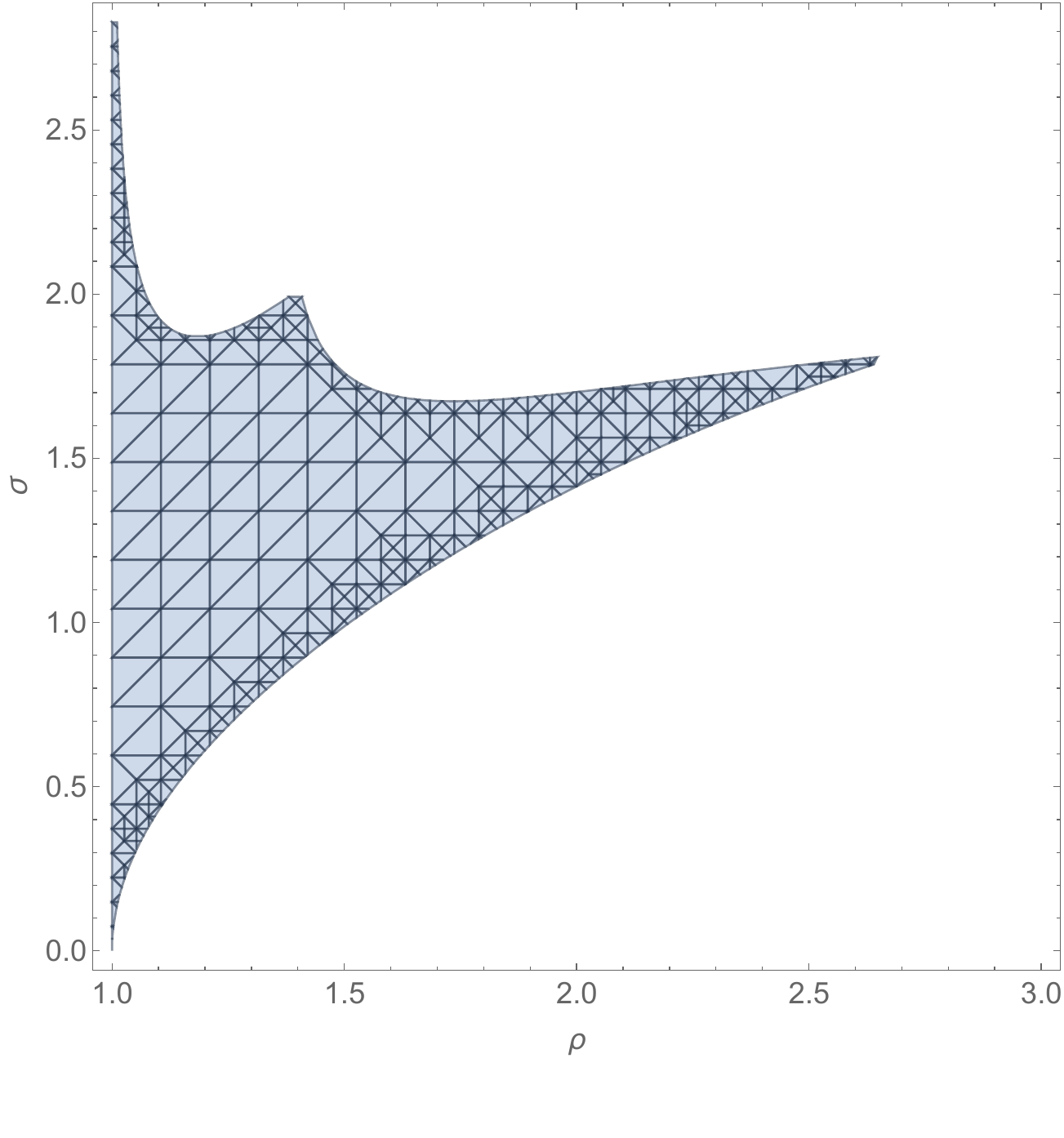}
\caption{On the left graph we have drawn the portion of the $(\rho,\s)$ plane for which the masses of the coupled fluctuations $\d z$ and $\d \tilde \psi$ are above the B-F bound, always subject to the condition \eqref{boundss}. As in figure \ref{figg-2}, $\s \in [0,2 \sqrt{2}]$ while $1< \rho \le 3$. On the right graph we have further imposed the constraint \eqref{co-2}. This is the final area in which $\rho$ and $\s$ can take values so that all masses are above the B-F bound.}
 \label{figg-2a}
 \end{figure}

\begin{figure}[h!]
% \centering
  %\includegraphics[width=0.55\textwidth]{ms1-bigger.pdf}
   %\includegraphics[width=0.55\textwidth]{ms2-bigger.pdf}
    \includegraphics[width=7.5cm,height=7cm]{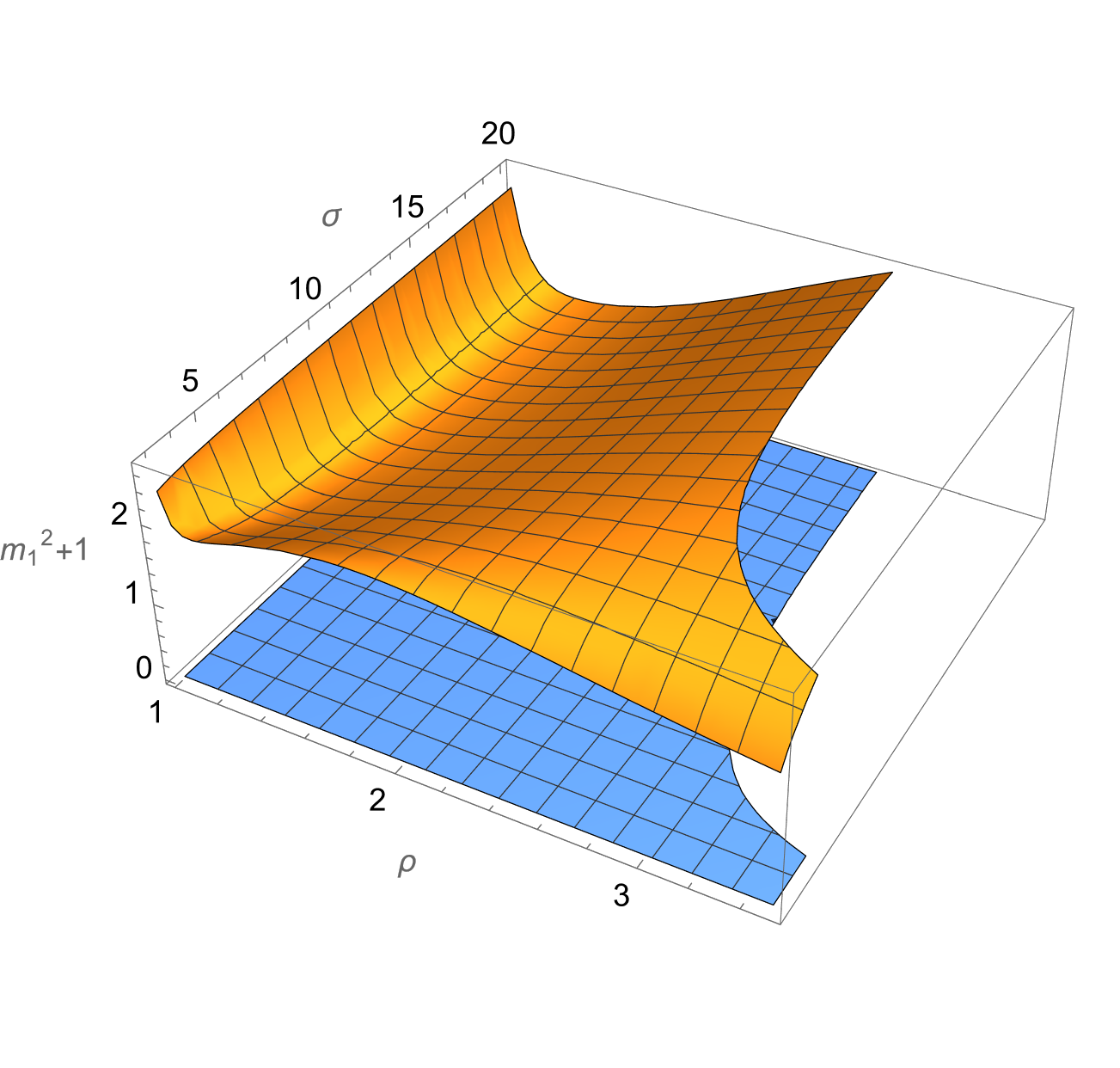}
    \includegraphics[width=7.5cm,height=7cm]{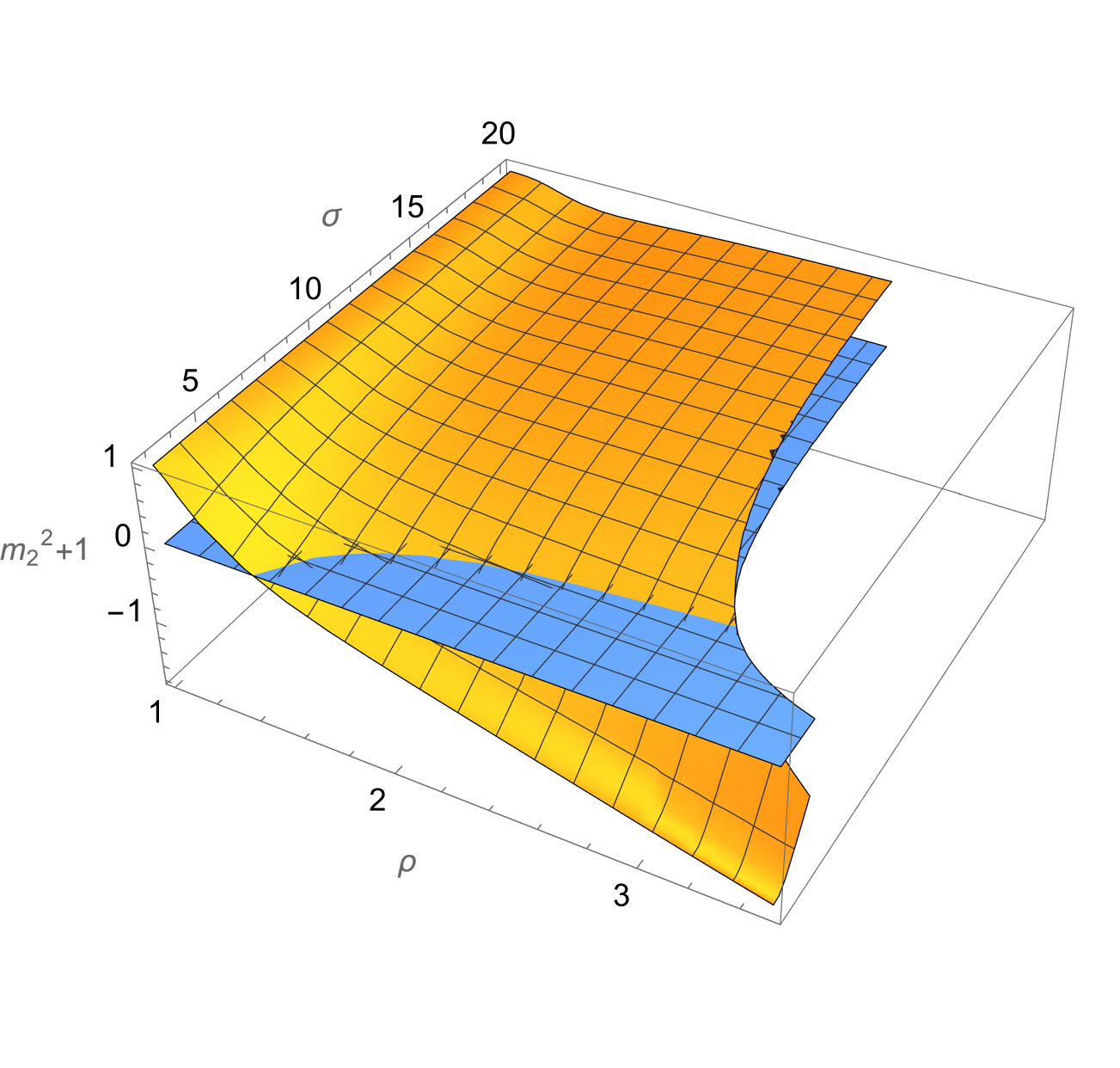}
 \caption{On the left graph we have drawn the function $m^2_1(\rho,\s)+1$ for the range of $\s\geq 2 \sqrt{2}$ subject to the condition \eqref{boundss}. We have allowed $\rho$ to be between $1$ and $3.5$. The blue plane is the zero of the vertical axis $z=0$. On the right graph we have drawn the function $m^2_2(\rho,\s)+1$ for the same values of $\rho$ and $\s$. }
 \label{figg-3}
 \end{figure}

 \begin{figure}[h!]
 %\centering
  %\includegraphics[width=0.45\textwidth]{ms2-bigger-region.pdf}
   %\includegraphics[width=0.45\textwidth]{ms2-bigger-region-restricted.pdf}
   \includegraphics[width=7.5cm,height=7cm]{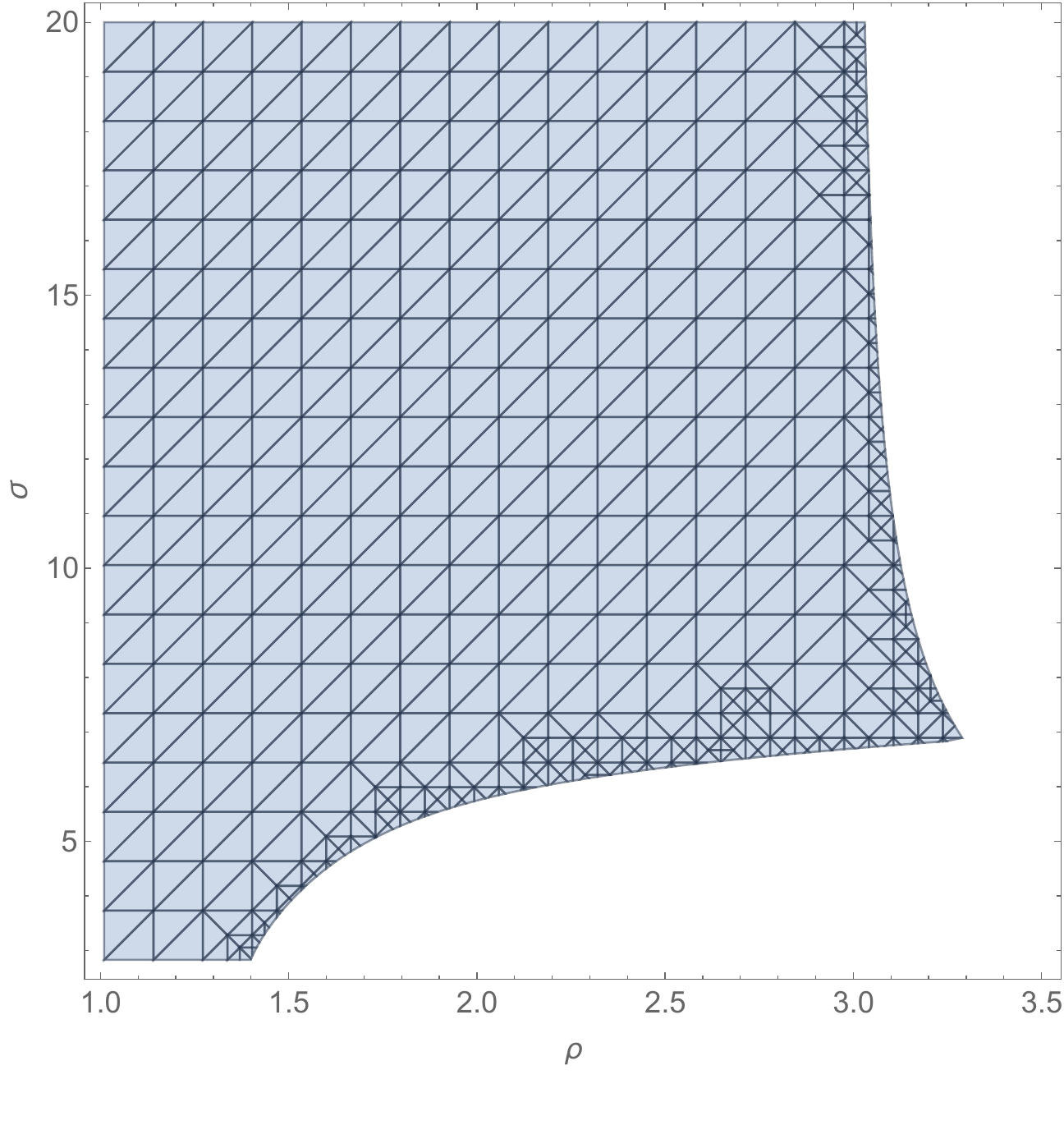}
    \includegraphics[width=7.5cm,height=7cm]{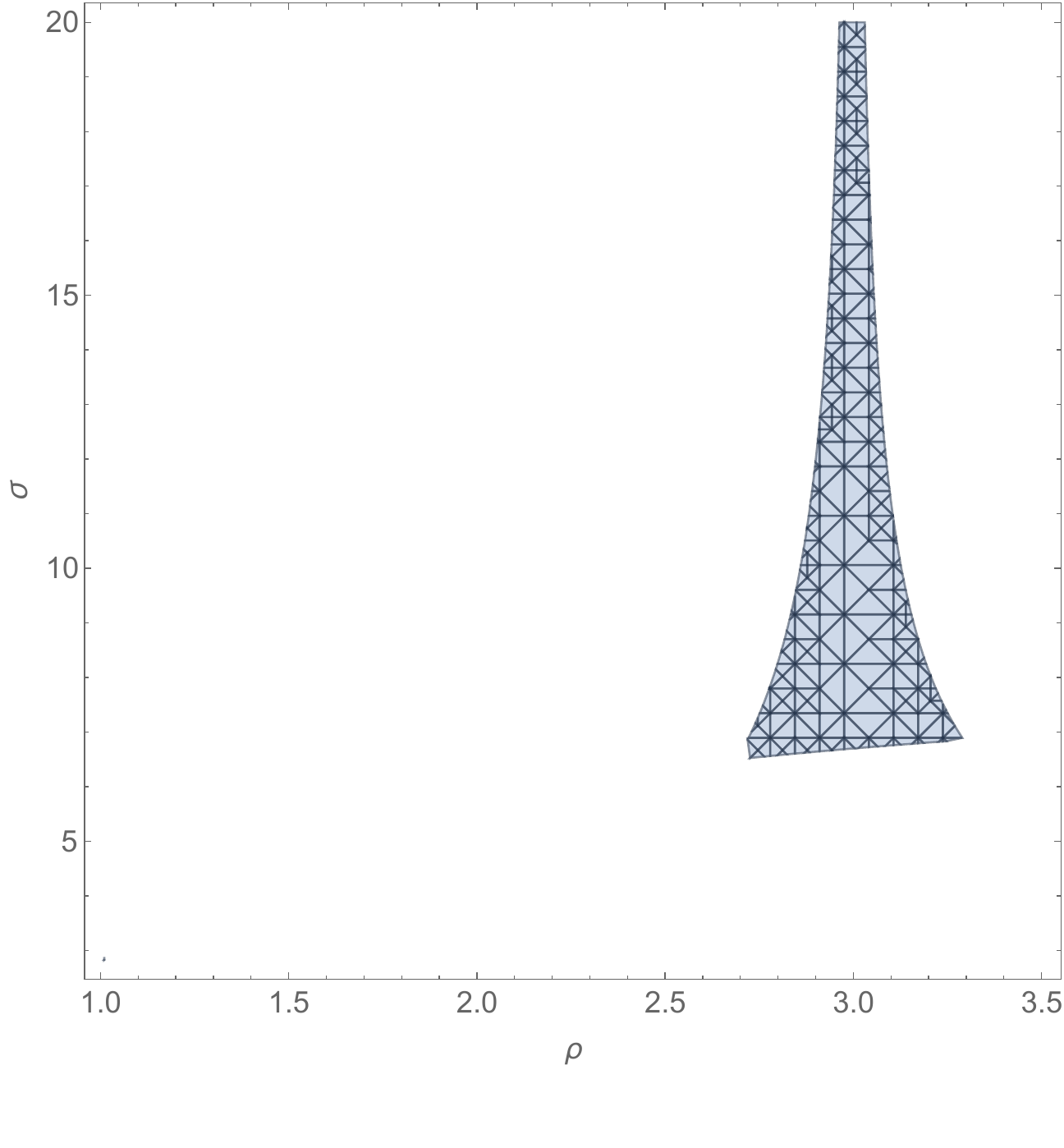}
\caption{On the left graph we have drawn the portion of the $(\rho,\s)$ plane for which the masses of the coupled fluctuations $\d z$ and $\d \tilde \psi$ are above the B-F bound, always subject to the condition \eqref{boundss}. As in figure \ref{figg-3}, $\s \geq 2 \sqrt{2}$ and $1 < \rho \le 3.5$. On the right graph we have further imposed the constraint \eqref{co-2}. This is the final area in which $\rho$ and $\s$ can take values so that all masses are above the B-F bound. }
 \label{figg-3a}
 \end{figure}

Finally, we consider the equations for the fluctuations of $ \d \tilde \psi(s)$ and $\d z(s)$. These equations are coupled and take the following form
\begin{eqnarray}\label{dz-dpsi}
&&\a_1(\rho,\s)\, \d z''(s)+\a_2(\rho,\s)\, \d z(s)+\a_3(\rho,\s) \,\d \tilde \psi(s)=0 \nonumber \\
&&\b_1(\rho,\s)\, \d \tilde \psi''(s)+\b_2(\rho,\s) \, \d \tilde \psi(s)+\b_3(\rho,\s)\,  \d z(s)=0,
\end{eqnarray}
where the functions $\a_i$ and  $\b_i$, $i=1,2,3$ are lengthy expressions 
of $\rho$ and $\sigma$ whose explicit form is not illuminating and will not be presented.
%given in the appendix. 
By making an ansatz of the form $\d z=A e^{i m s}$ and  $\d \tilde \psi=B e^{i m s}$ we arrive to the following system of equations
\begin{eqnarray}\label{masses-0}
(\a_2-\a_1 m^2)A+ \a_3 B=0 \quad {\rm and} \quad \b_3 A+(\b_2-\b_1 m^2)B=0 \, .
\end{eqnarray}
The homogeneous system in \eqref{masses-0} has a non-zero solution when its determinant is zero, i.e. when  $(\a_2-\a_1 m^2)(\b_2-\b_1 m^2)-\a_3 \b_3=0$. Solving this equation gives us the eigenvalues for the masses, which read
\begin{eqnarray}\label{masses}
&&m^2_1(\rho,\s)=\frac{\a_1\b_2+\a_2\b_1+\sqrt{(\a_2\b_1-\a_1\b_2)^2+4 \a_1\a_3\b_1 \b_3}}{2 \a_1\b_1} \nonumber \\[5pt]
&&m^2_2(\rho,\s)=\frac{\a_1\b_2+\a_2\b_1-\sqrt{(\a_2\b_1-\a_1\b_2)^2+4 \a_1\a_3\b_1 \b_3}}{2 \a_1\b_1}.
\end{eqnarray}
In figure \ref{figg-2} we draw the masses as functions of $\rho$ and $\s$ for the allowed parametric space as this is given in \eqref{boundss}. In particular,  in figure \ref{figg-2} we examine the case in which 
$0 \leq\s \leq2\sqrt{2}$. On the left graph the vertical axis is $m^2_1+1$ while on the right graph the vertical axis is $m^2_2+1$. In the first case, i.e. for $m^2_1+1$, we see that the corresponding surface is always above zero for all values of $\rho$ and $\s$ allowed by \eqref{boundss}, so the B-F bound is satisfied. For the second mass eigenvalue one can see that the B-F bound is satisfied only for the orange surface which is above the blue plane which is sitting at the zero of the vertical axis $z=0$. The projection of this surface on the $(\rho,\s)$ plane is depicted in the left of figure \ref{figg-2a}. Had the constraint from \eqref{co-2} not been present, this area would have given the allowed values for the pairs  $(\rho,\s)$. By further imposing the constraint \eqref{co-2} the allowed portion of the $(\rho,\s)$ plane is restricted a bit. The final allowed area in which $\rho$ and  $\s$ can take values is depicted in the right of figure  \ref{figg-2a}. As long as the parameters $\rho$ and  $\s$ belong to this range all masses satisfy the B-F bound and our solution is stable.

The analogous graphs for the case in which $\s>2\sqrt{2}$ and $1< \rho \le 5$ are shown in figure \ref{figg-3}. As in the previous case the mass of the first eigenvalue is always above the B-F bound (see graph in the left panel of figure \ref{figg-3}). The requirement that the second mass eigenvalue is above the B-F bound restricts the allowed values of $\rho$ and $\s$ as shown in the right graph of figure \ref{figg-3}, as well as on the left graph of \ref{figg-3a}. As above, the imposition of the constraint  \eqref{co-2} eliminates most of the allowed space for $\rho$ and $\s$. What remains is shown in the right graph of  figure \ref{figg-3a}. 
To recapitulate, the allowed values of $\rho$ and $\s$ are shown in the right graphs of figures \ref{figg-2a} and \ref{figg-3a}.

We conclude that the restrictions originating from the B-F bound impose further constraints other than those imposed from the equations of motion of the brane  \eqref{boundss}. The preceding analysis specifies the regions, right graphs in figures \ref{figg-2a} and \ref{figg-3a}, for which our D5 probe brane has no tachyonic instabilities.

\section{D5-D7 system and anomaly inflow}\label{D5-D7}

The careful reader may have noticed that, up to this point, we have ignored an essential feature of our solution, namely the fact that our D5 probe brane has boundaries. Indeed, since the value of $\rho$, in the equation  $\psi=\rho \, \tilde \gamma+\phi_0$ relating the boundary angle $\psi$ to the internal angle $\tilde \gamma$ ,  is not an integer the brane does not close on itself but it acquires, instead, 
a 5-dimensional boundary parametrized by $\{x_0,x_1,r,\b,\g\}$ which consists of two manifolds, that is $\partial M_{D_5}=\{x_0,x_1,r,\b,\g,\psi=\phi_0\}\bigcup\{x_0,x_1,r,\b,\g,\psi=2\pi\, \rho+\phi_0\}$.\footnote{In this section for convenience  we take the angle $\psi$ to be one of the D5 worldvolume coordinates instead of $\tilde\g$. However, we impose that $\tilde \g\in[0, 2 \pi)$ in agreement with section \ref{emb_ansatz}.} The existence of a boundary for the D5 brane will lead to inconsistencies because the D5 brane action will no longer be gauge invariant. To remedy this problem we place two D7 branes, one at $\psi=\phi_0$ and the other at $\psi=2\pi\, \rho+\phi_0$, on which our D5 brane can end. Then the anomalous terms of the D5 brane action will  be canceled through the anomaly inflow mechanism from the D7 branes. The precise way this cancellation happens will be presented in what follows.

Before that, let us describe the D7 branes that we will be using. We choose the embedding coordinates of the D7 brane to be $\{x_0,x_1,r,z,\tilde \b,\tilde\g,\b,\g\}$.
Then one can find the following solution to the equations of motion derived from the action
\begin{equation}
\label{D7-Lor}
S_{D7}=- \frac{T_7}{g_s}\Bigg\{\int d^8 \zeta \sqrt{-{\rm det}\, \mathcal P [g+2 \pi \a' F]}-
\int \mathcal P [ e^{B+2 \pi \a' F}\wedge C_4]\Bigg\} .
\end{equation}
The solution reads\footnote{Actually, one can show the fluctuations of the coordinate $\tilde\psi$ have mass above the BF bound and as a result the solution is free of tachyonic instabilities if $\tilde\psi_0\in [0,\cos ^{-1}\left(4 \sqrt{\frac{3}{73}}\right)]\cup [\cos ^{-1}\left(\frac{5}{\sqrt{73}}\right),\frac{\pi}{2}]\approx [0,0.625]\cup [0.946,\frac{\pi}{2}]$. For the special case $\tilde\psi_0=\frac{\pi}{4}$ the instability was shown in \cite{Bergman:2010gm}.} 
\begin{eqnarray}\label{sol-D7}
&\psi=\psi_0, \quad \tilde\psi=\tilde\psi_0\quad {\rm with } \quad   A^{(D7)} = \frac{\pm \sin{\tilde\psi_0}\sqrt{\cos{2\tilde\psi_0}}}{2 \, \pi \, \alpha'} \, \cos \tilde \beta \, d\tilde\gamma  , \quad{\rm for }\quad  
\frac{\pi}{4}\geq \tilde\psi_0\ge 0 \\
&\psi=\psi_0, \quad \tilde\psi=\tilde\psi_0\quad {\rm with } \quad   A^{(D7)} = \frac{\pm \cos{\tilde\psi_0}\sqrt{-\cos{2\tilde\psi_0}}}{2 \, \pi \, \alpha'} \, \cos \beta \, d\gamma  , \quad{\rm for }\quad  
\frac{\pi}{2}\geq \tilde\psi_0>\frac{\pi}{4}.\nonumber
\end{eqnarray}
In \eqref{sol-D7} $\psi_0= \phi_0$ or $\psi_0= 2 \pi \rho+\phi_0$.
Notice that for each range of the internal angle $\tilde\psi_0$ the worldvolume field is turned on along different directions. The above D7 solution is chosen to sit at a fixed value of the coordinate 
$\psi$ so that the D5 brane can have its boundary at a fixed value of $\psi$.
 At this point, let us mention that the solution \eqref{sol-D7} is a special case of the solution studied in  \cite{Bergman:2010gm,Kristjansen:2012tn}. Indeed, if in the solution of \cite{Kristjansen:2012tn,Bergman:2010gm} one sets one of the two fluxes to zero, that is $k_1=0$ or  $k_2=0$, the parameter $\Lambda$ of \cite{Kristjansen:2012tn,Bergman:2010gm} becomes zero while the remaining non-zero flux become that of \eqref{sol-D7}. As we will see in what follows, the choice of this special D7 brane with one of the fluxes being zero has minimal consequences on the dual CFT.  

Our construction proceeds now as follows. We place two of the aforementioned D7 branes,  one at $\psi=\phi_0$ and the other at $\psi=2\pi\, \rho+\phi_0$, on which our D5 brane terminates. 
As mentioned above, the D5 ends on a 5-dimensional sub-manifold of the D7 parametrised by $\{x_0,x_1,r,\b,\g\}$. As it commonly happens, the D7 brane sees this boundary of the D5 as a magnetic monopole that violates the Bianchi identity of the D7 brane, see for example \cite{Bachas:1998rg}
\be\label{D7-monopole}
dF^{(D7)}=2 \pi\, g_{D7}\,\d(z-\s r)\,  \d(\tilde \b-\frac{\pi}{2}) \ \d(\tilde \g)\, dz \wedge d\tilde \b \wedge d\tilde \g \, .
\ee
Notice that the role of the 3 delta functions is to localise the magnetic monopole on the sub-manifold of the D7 brane which is the boundary of the D5 probe brane.
We now consider the gauge variations of the D5 and D7 brane actions. To this end, we employ the democratic formulation of the type IIB supergravity in which the RR-fields $C_p,\,\,p=0,2,4,6,8$
 transform under gauge transformations as
 \be\label{gauge-trans}
 \d C_p=d \L_{p-1}+H\wedge \L_{p-3}, 
 \ee
 where $H=dB$ is the field strength of the NS B-field and $\L_{p-1}$ are the parameters of the corresponding gauge transformations.\footnote{When $p-3<0$, i.e. for $p=0,2$ the second term on the right hand side of \eqref{gauge-trans} is absent.}
 
 Let us now focus on the variation of the term in the Wess-Zumino (WZ) action  of the D5 brane which depends on $C_6$. It is given by
 \begin{IEEEeqnarray}{l}\label{D5-C6}
 \d S_{D5}^{(C_6)}= T_5 \int_{M_{D5}} \d C_6= T_5 \int_{M_{D5}} d \L_5+H\wedge  \L_3 \approx  T_5 \int_{\partial M_{D5}}  \L_5. 
 \end{IEEEeqnarray}
 One can check that the bulk second term involving the gauge parameter $\L_3$ cancels against an opposite term coming from the variation of the WZ term that is proportional to 
 $C_4\wedge (B+2 \pi \a' F)$. The total derivative term is that it will contribute to the anomaly. This term is zero when the D5 brane has no boundary. In our case the D5 brane has a boundary and the anomaly of \eqref{D5-C6} should be cancelled against the gauge variation of the term in the Wess-Zumino (WZ) action  of the D7 brane which depends on $C_6$, namely
 %\begin{IEEEeqnarray}{l}\label{D7-C6}
 \be\label{D7-C6}
\frac{ \d S_{D7}^{(C_6)}}{2 \pi \a'}= T_7 \,\d \int_{M_{D7}} C_6\wedge F^{(D7)}= T_7 \int_{M_{D7}} d \L_5 \wedge F^{(D7)}+\cdots \approx  T_7 \int_{ M_{D7}}  \L_5 \wedge dF^{(D7)},
 %\end{IEEEeqnarray}
 \ee
 where the dots denote terms  which do not contribute to the anomaly. Taking now into account the violation of the Bianchi identity of the D7 brane \eqref{D7-monopole} the variation becomes
 \be\label{D7-C6-1}
\frac{ \d S_{D7}^{(C_6)}}{2 \pi \a'}= 2 \pi g_{D7}\,T_7 \int_{\partial M_{D5}} \L_5.
\ee
Notice that the integral in \eqref{D7-C6-1} is over the boundary of the D5 since the 3 delta functions in the \eqref{D7-monopole} localise the integral of  \eqref{D7-C6}.
 The strength of the magnetic monopole $g_{D7}$ is determined by the requirement that the sum of the variations \eqref{D5-C6} and \eqref{D7-C6-1} is zero.
 By taking into account that %the flux {\color{red}due to the monopole} through the $S^2$ parametrised by $(\tilde \b,\tilde\g)$ is $\int_{S^2_{(\tilde \b,\tilde\g)}} F^{(D7)}=2 \pi \, g_{D7}$ and 
 that 
 the tension of a $Dp$ brane is given by $T_p=\frac{1}{(2 \pi)^p g_s\a'^{\frac{p+1}{2}}}$ we obtain the anomaly cancellation condition to be 
 \be\label{C6-cancel}
 T_5+T_7 \,2 \pi \a' 2 \pi g_{D7}=0 \Rightarrow g_{D7}=-1.
 \ee
 So the strength of the magnetic monopole that the D7 brane feels, due to the fact that the D5 brane ends on it, is one. At this point, let us mention that due to the fact that the two boundaries of the D5 brane, one at $\psi=\phi_0$ and the other at $\psi=2 \pi \rho+\phi_0$, have opposite orientations the strength of the magnetic monopole that the D7 sees is $g_{D7}=-1$ and $g_{D7}=+1$, respectively.
 
 We now turn to the anomaly cancellation of the terms that are proportional to the gauge parameter $\L_3$.
 The contribution of the D7 brane to this anomaly is given by
 \begin{IEEEeqnarray}{l}\label{D7-C4}
 \frac{ \d S_{D7}^{(C_4)}}{(2 \pi \a')^2}= \frac{T_7}{2} \,\d \int_{M_{D7}} C_4\wedge F^{(D7)}\wedge F^{(D7)}= \frac{T_7}{2} \int_{M_{D7}} d \L_3 \wedge F^{(D7)} \wedge F^{(D7)}+\cdots \approx \nonumber\\
  T_7 \int_{ M_{D7}}  \L_3\wedge F^{(D7)} \wedge dF^{(D7)}= T_7 \int_{ \partial M_{D5}}  \L_3\wedge F_{S^2_{( \b,\g)}} ^{(D7)} 2 \pi g_{D7},
  \end{IEEEeqnarray}
 where we have used \eqref{D7-monopole} and $ F_{S^2_{( \b,\g)}} ^{(D7)}$ is the part of the 2-form worldvolume which appears in the D7 brane solution \eqref{sol-D7}. 
 Finally, by taking into account that $g_{D7}=-1$ and that the flux through the $S^2_{( \b,\g)}$ should be an integer $m_2$, i.e. $\int_{S^2_{( \b,\g)}} F_{S^2_{( \b,\g)}}^{(D7)}=2 \pi \, m_{2}$ we obtain
 \begin{IEEEeqnarray}{l}\label{D7-C4-1}
 \frac{ \d S_{D7}^{(C_4)}}{(2 \pi \a')^2}= -T_7 2 \pi  \frac{m_{2}}{2}\int_{ \partial M_{D5}} \L_3\wedge d\Omega_{S^2_{( \b,\g)}}
 \quad {\rm with} \quad 
 d\Omega_{S^2_{( \b,\g)}} = \sin \beta \, d \b \wedge d\g \, . 
 \end{IEEEeqnarray}
 
As it will become apparent in a while, in order to cancel the gauge anomaly that is proportional to $\L_3$, one should insert on the boundary of the worldvolume of the D5 brane, $ \partial M_{D5}$, a magnetic monopole of the form
\be\label{D5-monopole}
dF^{(D5)}=2 \pi\, g_{D5}\,\d(\psi-\psi_0)\,  d\psi \wedge d\Omega_{S^2_{( \b,\g)}}, \quad \psi_0=\phi_0 \lor \psi_0=2 \pi \rho+\phi_0\, .
\ee
At this point let us mention that it is perfectly consistent to put the aforementioned magnetic monopole on the boundary of the D5 brane. It accounts for the different fluxes, through the common $S^2_{\b,\g}$, of the  D5 and D7 branes. Namely, it is the sum of the flux of the monopole plus that of the D5 brane that flows into the D7 brane. As we will see shortly, this flux conservation is ultimately connected to the cancellation of the gauge anomaly.

We are now in position to evaluate the gauge variation of the D5 brane WZ term proportional to $C_4$. It reads
\begin{IEEEeqnarray}{l}\label{D5-C4}
\frac{ \d S_{D5}^{(C_4)}}{2 \pi \a'}= T_5 \,\d \int_{M_{D5}} C_4\wedge F^{(D5)}= T_5 \int_{M_{D5}} d \L_3 \wedge F^{(D5)}+\cdots \approx  T_5 \int_{ M_{D5}}  \L_3 \wedge dF^{(D5)}+\nonumber \\
T_5 \int_{ M_{D5}} d( \L_3 \wedge F^{(D5)})\overset{\eqref{D5-monopole}}{=}T_5 2 \pi\,g_{D5}\int_{\partial M_{D5}}  \L_3 \wedge 
d\Omega_{S^2_{( \b,\g)}}+T_5 \, \int_{\partial M_{D5}}  \L_3 \wedge F_{S^2_{( \b,\g)}} ^{(D5)}
\end{IEEEeqnarray}
 where $F_{S^2_{( \b,\g)}} ^{(D5)}$ is the part of the worldvolume 2-form of the D5 brane derived from  \eqref{A}.\footnote{The part of the D5 brane 2-form originating from the integration of the D5 monopole of \eqref{D5-monopole} will not contribute in the last term of \eqref{D5-C4}. We will argue that this is, indeed, the case 
 in appendix \ref{Appendix:details_monopoles}.}
 Next we demand that the flux through the two-spheres should be integer multiples of $2 \pi$, namely that $\int_{S^2_{(\b,\g)}}F_{S^2_{( \b,\g)}} ^{(D5)}=2 \pi m_1\equiv2 \pi k $ and that from \eqref{D5-monopole}
 the non-closed contribution to the flux through $S^2_{( \b,\g)}$ is $\int_{S^2_{(\b,\g)}}F_{monopole} ^{(D5)}\equiv2 \pi m_{D5}=2 \pi g_{D5} 4 \pi\Rightarrow 2 \pi g_{D5}=\frac{m_{D5}}{2}$. Putting everything together and adding up 
 \eqref{D7-C4-1} and \eqref{D5-C4} one sees that the gauge anomaly vanishes, as long as
  \be\label{C4-total}
  - 2\pi m_2 \,T_7 \,2 \pi \a'+T_5\, m_{D5}+T_5 \,m_1=0\quad \Longrightarrow \quad m_{D5}=m_2-m_1.
  \ee
Of course, due to the opposite orientations of the two boundaries of the D5 brane which sit one at $\psi=\phi_0$ and the other at $\psi=2 \pi \rho+\phi_0$, the two monopoles inserted should have opposite strength, that is one boumdary should carry a monopole $m_{D5}=m_2-m_1$ and the other boundary an anti-monopole $m_{D5}=-(m_2-m_1)$.
 
Let us now interpret  \eqref{C4-total}. The first thing to notice is that the D5 and the D7 brane share a common $S^2_{(\b,\g)}$ along the directions $\b$ and $\g$. This implies that the flux through this two-sphere should be equal in both the D5 and the D7 brane solutions. But apparently this does not happen as can be easily seen from \eqref{A} and \eqref{sol-D7}. This fact will lead
 to an anomaly unless a magnetic monopole -to be precise a monopole and an anti-monopole- of the form \eqref{D5-monopole} is placed on each of the boundaries of the D5 brane. The role of the monopoles is to absorb part of the flux of the D5 brane.  The strength of the monopoles is such that  the flux is conserved  through the two-sphere $S^2_{(\b,\g)}$. In this way the gauge anomaly becomes zero.

 In conclusion, in this section we have seen that the gauge anomaly which arises due to the fact that the D5 brane has boundary is cancelled through anomaly inflow from a set of two D7 branes 
 that we have put in the geometry and on which the D5 brane terminates. The resulting system of the two D7 and the single D5 brane has an additional gauge anomaly originating from the variation of the 4-form $C_4$. This anomaly is cancelled by inserting two magnetic monopoles of opposite strength. One of them sits at the boundary $\psi=\phi_0$ of the D5 brane $\partial M_{D5}$ while the other at the boundary $\psi=2 \pi \rho+\phi_0$. The strength of the monopoles is such that the flux, 
 through the common two-sphere $S^2_{(\b,\g)}$ of the 
 D5 and the D7 solutions, is conserved.
 As another comment let us mention that the one-loop anomaly originating from chiral fields is zero in our case because the boundary of the D5 brane is five (odd) dimensional and as a consequence does not admit this kind of anomaly \cite{Bilal:2004xt}. Furthermore, the phenomenon of anomalous creation of branes discussed in \cite{Bachas:1997ui} can not occur in our construction because the D5-D7 intersection is 5-dimensional while the D7-D7 intersection is 7-dimensional. For the anomalous creation to take place the dimension of the intersection should be 2 mod 4 \cite{Bachas:1997ui}.

A final comment concerning the fate of the D7  branes at the two endpoints $\tilde\psi_0=0$ and $\tilde\psi_0=\frac{\pi}{2}$ is in order. At each of these points one of the two-spheres around which the D7 branes wrap shrinks to zero while at the same time the worldvolume gauge field goes to zero in a certain rate and as a result their on-shell action is zero.  In any case, the perspective we take here is that they decouple from the D5 when either of the  limits is reached. This is consistent with the fact that the D7 branes are no longer necessary for the consistency of the D5 solution. We believe that at these limits the system of decoupled D5 and D7 branes will be energetically favoured compared to the coupled D5-D7 system.
 Thus, at these limits, the D7 branes can somehow be ignored.

\section{Conformal Field Theory dual}\label{dual}

In this section, we will discuss the details of the defect CFTs which are dual to our D5-brane solutions. These CFTs will be realised as 
classical solutions of the  ${\cal N}=4$ SYM equations of motion that describe the two dimensional interface generated by the D5
brane. 

Our starting point is the ${\cal N}=4$ SYM Lagrangian density in the mostly plus Lorentzian signature which reads
\begin{equation} \label{LagrangianSYM}
\LL_{\N = 4} = \frac{2}{g_{\text{\scalebox{.8}{YM}}}^2} \text{tr}\bigg\{-\frac{1}{4} F_{\mu\nu} F^{\mu\nu} - \frac{1}{2} \left(D_{\mu}\varphi_i\right)^2 + \frac{i}{2}\,\bar{\psi}\slashed{D}\,\psi +\frac{1}{2}\,\bar{\psi}\,\Gamma^{i+3}\, \left[\varphi_i, \psi \right]+ \frac{1}{4}\left[\varphi_i,\varphi_j\right]^2\bigg\} 
\end{equation}
where, as usual, $\bar{\psi}_{\alpha} \equiv \psi_{\alpha}^{\dagger} \Gamma^0$, $\slashed{D} \equiv \Gamma^{\mu}D_{\mu}$ and the gauge field strength and covariant derivatives are defined as 
\begin{IEEEeqnarray}{l}
F_{\mu\nu} \equiv \partial_{\mu}A_{\nu} - \partial_{\nu}A_{\mu} - i \left[A_{\mu},A_{\nu}\right], \quad D_{\mu}f \equiv \partial_{\mu}f - i \left[A_{\mu},f\right] \, .  \label{CovariantDerivatives}
\end{IEEEeqnarray}
Here $\Gamma^{M}$ and $\psi$ denote the 10-dimensional gamma matrices and fermion fields, respectively. The matrices $\Gamma^M$ obey the 10-dimensional Clifford algebra, 
%namely $\{\Gamma_M,\Gamma_N \}=-2\, \eta_{MN} {\mathbf 1}_{10}$, 
with the capital index $M$ being $M=(\mu,i+3),\, \mu=0,1,2,3, \,i= 1,2,3,4,5,6$. Furthermore, $\varphi_i$ denote the six real scalar fields of $\mathcal{N}=4$ SYM. All fields are taken to transform in the adjoint representation of $\mathfrak{u}(N)$.

From \eqref{LagrangianSYM}, one may derive the equations of motion for the bosonic fields which read
\begin{IEEEeqnarray}{c}\label{eoms}
D^{\mu}F_{\mu\nu} = i\left[D_{\nu}\varphi_i,\varphi_i\right], \qquad D^{\mu}D_{\mu}\varphi_i = \left[\varphi_j,\left[\varphi_j,\varphi_i\right]\right]. %\\
%i\slashed{D}\psi_{\alpha} = \sum_{i=1}^{3}G_{\alpha\beta}^{i}\left[\psi_{\beta},\varphi_i\right] + \sum_{i=4}^{6} G_{\alpha\beta}^{i}\gamma_5\left[\psi_{\beta},\varphi_i\right].
\end{IEEEeqnarray}
In what follows, all fermionic fields will consistently be set to zero. As a result, their equations of motion are trivially satisfied. 

\subsection{The D7 contribution}\label{D7-contribution}
Let us start by considering the effect of the D7 branes on the field theory dual. These  branes will introduce a co-dimension 1 defect\footnote{For co-dimension 1 defects see the recent review \cite{Linardopoulos25a} and references therein.} sitting at the boundary value $\psi=\psi_0$, with  $\psi_0=\phi_0$ or $\psi_0=2 \pi \rho+\phi_0$. The fact that one of the fluxes is zero (see eq. \eqref{sol-D7}) implies that the holographic one-point function of BPS operators will be zero.  The vanishing of the one-point functions implies in turn that the expectation values for all six bulk scalars should be set to zero. This can be seen alternatively as follows. For generic non-zero fluxes $k_1\neq0\neq k_2$, the field theory dual of the D7 branes of \cite{Kristjansen:2012tn} is given by 
\begin{align}  \label{classicalsolution1}
\varphi_i(z)&=-\frac{1}{z} (t_i^{k_1} \otimes 1_{k_2\times k_2})\oplus
0_{(N-k_1k_2)\times (N-k_1 k_2)},
\hspace{0.7cm} \mbox{for}\hspace{0.3cm} i=1,2,3, \\
\label{classicalsolution2}
\varphi_i(z)&=-\frac{1}{z} (1_{k_1\times k_1}\otimes t_i^{k_2})\oplus
0_{(N-k_1k_2)\times (N-k_1 k_2)},
\hspace{0.7cm} \mbox{for}\hspace{0.3cm} i=4,5,6.
\end{align}
Here $z$ is the distance from the co-dimesion 1 defect created by the D7 brane. Furthermore, $t_i^{k_1}$ and $t_i^{k_2}$
 are generators of the $k_1$-dimensional and $k_2$-dimensional irreducible
representation of $SU(2)$, respectively, while
$1_{k_1\times k_1}$ and $1_{k_2\times k_2}$ denote unit matrices of dimension 
$k_1\times k_1$ and $k_2\times k_2$. In the case where $k_1=0\lor k_2=0$ the expressions \eqref{classicalsolution1} and \eqref{classicalsolution2} become meaningless since there are no matrices of zero dimension. The correct way to interpret this is to say that the corresponding vevs become zero, namely 
\be\label{sol12}
\boxed{\varphi_i(z)=0_{N\times N},
\hspace{0.7cm} \mbox{for}\hspace{0.3cm} i=1,2,3,4,5,6\, .}
\ee
The last equation implies that the weak coupling one-point function of any CPO will be zero $\langle {\mathcal O}_\D\rangle=0 $. This prediction is in agreement with the strong coupling calculation of CPOs \cite{Kristjansen:2012tn} which gives  $\langle {\mathcal O}_\D\rangle\sim k_1 k_2\overset{k_1=0}{=}0  $. Some more details of the field theory dual of the D7 branes can be found in appendix \ref{Appendix:details_FT-dual}.

In conclusion, we see that the effect of the D7 branes is minimal in the sense that they do not introduce any expectation value for the bulk scalars $\varphi_i$ but only for the scalar fields of the  hypermultiplets living on the defect (see appendix \ref{Appendix:details_FT-dual}).

\subsection{The combined D5-D7 contribution}\label{D7-contr}

The aim of the rest of this section is to motivate the field theory dual of the D5-brane solutions presented in section \ref{emb_ansatz}.
These probe D5-brane solutions provide the bulk description of a novel class of non-supersymmetric defect operators
of   ${\cal N}=4$ SYM which are supported on the surface $\S= \mathbb{R}^{(1,1)}$, parametrised by $(x_0,x_1)$ in our construction.
These nonlocal surface operators, which we will denote by ${\cal O}_\Sigma$, will induce  a  co-dimension 2
singularity for the classical fields of ${\cal N}=4$ SYM. 
Similarly to what happens for the half-BPS surface defects of \cite{Gukov:2006jk}, 
the singularity in the gauge field produced by the surface operator would be that of a non-abelian vortex configuration which in our case  takes the form 
\be\label{A-sing}
A= \left(\begin{array}{cc} a_1 \otimes I_{(N_1+k) \times (N_1+k)} &\,\,\,\, 0_{(N_1+k) \times \left(N -N_1-k \right)} \\ 0_{\left(N - N_1-k\right)\times (N_1+k)} &\,\,\,\,   0_{\left(N -N_1- k\right)\times \left(N-N_1 - k\right)} \end{array}\right)_{N\times N}\, d\psi,
\ee
where $N_1$ is the number of the D5 branes and $k$ is the integer flux through the $S^2_{(\b,\g)}$ in the D5 solution (see \eqref{kappa-1}).
Notice that since we are in the probe approximation we have set the lower right part of $A$ to zero.

In \eqref{A-sing} we have considered a gauge singularity which breaks the $G=U(N)$ gauge group down to $U(N_1+k)\times U(N-N_1-k)$.  We remind the reader that $\psi$ is the polar angle in the plane perpendicular to the surface of the defect $\S$, while $a_1$ characterises the surface operator. The  gauge field $A$ in \eqref{A-sing}  satisfies the corresponding  equation of motion (left equation of \eqref{eoms}) away from the location of the surface operator given that the commutator in the right hand side is zero. We will see that this will be, indeed, the case.
The corresponding field strength is then given by 
\be
F=dA=2 \pi \left(\begin{array}{cc} a_1 \otimes I_{(N_1+k) \times (N_1+k)} &\,\,\,\, 0_{(N_1+k) \times \left(N -N_1-k \right)} \\ 0_{\left(N - N_1-k\right)\times (N_1+k)} &\,\,\,\,  0_{\left(N -N_1- k\right)\times \left(N-N_1 - k\right)} \end{array}\right)\d_\S  \,\,\,  {\rm where}  \,\,\,  \d_\S=d(d\psi)
\ee
and as mentioned above, satisfies the eom for $A$. Here, $\d_\S=d(d\psi)$ is a two-form delta function supported on the surface of the operator.

Given that the existence of the gauge field in \eqref{A-sing} breaks the gauge group from $U(N)$ to $U(N_1+k)\times U(N-N_1-k)$ 
there is an additional freedom in the definition of ${\cal O}_\Sigma$ consistent with all the symmetries, namely one can turn on two dimensional $\theta$-angles for the
unbroken $U(1)$s along the surface of the defect $\S$ \cite{Gukov:2006jk,Gomis:2007fi,Drukker:2008wr}. In particular one can insert in the ${\cal N}=4$ SYM  path integral the following operator which also characterises the defect
\be\label{insertion}
e^{i \,  \eta_1 \int_\S Tr F^{(1)}}
\ee
where $F^{(1)}$ is the field-strength  and $\eta_1$  the two dimensional $\theta$-angle \cite{Gukov:2006jk,Gomis:2007fi,Drukker:2008wr}. %In what follows, we will choose $\eta_2=0$. The reason for this choice will be clear in a while.
Thus, there is an additional $N\times N$ matrix which characterises the surface operator. It collects the  $\theta$-angles and commutes with the gauge field $A$ having the form
\be\label{eta-sing}
\eta=\left(\begin{array}{cc} \eta_1 \otimes I_{(N_1+k) \times (N_1+k)} &\,\,\,\, 0_{(N_1+k) \times \left(N -N_1-k \right)} \\ 0_{\left(N - N_1-k\right)\times (N_1+k)} &\,\,\,\,  0_{\left(N -N_1- k\right)\times \left(N-N_1 - k\right)} \end{array}\right)_{N\times N}\, .
\ee
Needless to say that the gauge field producing the field strengths $F^{(1)}$ is proportional to $\eta_1$.

In the case of ${\cal N}=4$ SYM, the surface operator induces a further singularity for the scalar fields along the defect $\S$. To determine which of the scalar fields acquire vacuum expectation values (vevs) we first notice that in the coordinate system \eqref{metric} the six coordinates of the $S^5$, $X_i,\,\,i=1,\cdots 6$ satisfy $X_1^2+X_2^2+X_3^2=\sin^2{\tilde \psi}$ and $X_4^2+X_5^2+X_6^2=\cos^2{\tilde \psi}$. 
More precisely, we have 
\begin{IEEEeqnarray}{l}\label{coord-sphere}
X_1=\cos{\tilde \b}\sin{\tilde \psi}, \quad X_2=\sin{\tilde \b}\cos{\tilde \g}\sin{\tilde \psi},  \quad X_3=\sin{\tilde \b}\sin{\tilde \g}\sin{\tilde \psi}\nonumber \\
X_4=\cos{ \b}\cos{\tilde \psi}, \quad  X_5=\sin{ \b}\cos{\g}\cos{\tilde \psi} , \quad X_6=\sin{ \b}\sin{ \g}\cos{\tilde \psi}\, .
\end{IEEEeqnarray}
%$X_1=\cos{\tilde \b}\sin{\tilde \psi}$, $X_2=\sin{\tilde \b}\cos{\tilde \g}\sin{\tilde \psi}$ and $X_3=\sin{\tilde \b}\sin{\tilde \g}\sin{\tilde \psi}$ and $X_4=\cos{ \b}\cos{\tilde \psi}$, $X_5=\sin{ \b}\cos{\g}\cos{\tilde \psi}$ and $X_6=\sin{ \b}\sin{ \g}\cos{\tilde \psi}$.

Then, the symmetries of the system allows one to make for the bulk scalars the following educated guess 
\begin{empheq}[box=\fbox]{align}\label{sol-2}
\varphi_{i+3}^{\text{cl}}\left(r'\right) &= \frac{1}{\sqrt{2}\,r'}   \cdot \left[\begin{array}{cc} \left(t_i\right)_{k\times k} & 0_{k\times \left(N - k\right)} \\ 0_{\left(N - k\right)\times k} & 0_{\left(N - k\right)\times \left(N - k\right)} \end{array}\right], \quad i=1,2,3 
\\[5pt]
\varphi_{i}^{\text{cl}}\left(r',\psi\right) &= \big(\d_{i2}\Re[ \frac{1}{z}C]+\d_{i3}\Im[ \frac{1}{z}C]\big) \sin{\tilde \psi}_0\, . \nonumber 
\end{empheq}
%\begin{IEEEeqnarray}{l}\label{sol-2}
%  \varphi_{i+3}^{\text{cl}}\left(r'\right) = \frac{1}{\sqrt{2}\,r'}   \cdot \left[\begin{array}{cc} \left(t_i\right)_{k\times k} & 0_{k\times \left(N - k\right)} \\ 0_{\left(N - k\right)\times k} & 0_{\left(N - k\right)\times \left(N - k\right)} \end{array}\right], \quad i=1,2,3\\
%\varphi_{i}^{\text{cl}}\left(r',\psi\right) = \big(\d_{i2}\Re[ \frac{1}{z}C]+\d_{i3}\Im[ \frac{1}{z}C]\big) \sin{\tilde \psi}_0\, . \nonumber 
%\end{IEEEeqnarray}
The second equation in \eqref{sol-2} can be alternatively written as
\begin{IEEEeqnarray}{l}\label{sol-2a}
% \varphi_{i+3}^{\text{cl}}\left(r'\right) = \frac{1}{\sqrt{2}\,r'} \cdot \left[\begin{array}{cc} \left(t_i\right)_{k\times k} & 0_{k\times \left(N - k\right)} \\ 0_{\left(N - k\right)\times k} & 0_{\left(N - k\right)\times \left(N - k\right)} \end{array}\right], \quad  
% i=1,2,3 \nonumber\\
\boxed{ \varphi_{1}^{\text{cl}}\left(r'\right) = 0,\quad \varphi_{2}^{\text{cl}}(r',\psi) +i \varphi_{3}^{\text{cl}} (r',\psi)=   \sin{\tilde \psi}_0\frac{1}{z}\,C\, ,}
\end{IEEEeqnarray}
where $r'$ is the radial distance from the co-dimension 2 defect $r'=\sqrt{x'^2_2+x'^2_3}$, $z=r' e^{i\psi}$ and $C$ is the $N\times N$ matrix
\be\label{C-matr}
\boxed{ C=\left(\begin{array}{cc} (\b_1+i \g_1) \otimes I_{(N_1+k)\times (N_1+k)} &\,\,\,\, 0_{(N_1+k)\times \left(N-N_1 - k\right)} \\ 0_{\left(N-N_1- k\right)\times (N_1+k)} &\,\,\,\,  0_{\left(N-N_1 - k\right)\times \left(N-N_1 - k\right)} \end{array}\right)_{N\times N}\, .  }
\ee
Notice that we have set the vev of the scalar $ \varphi_{1}^{\text{cl}}=0$ since the corresponding coordinate  in the D5 brane solution is zero $X_1\sim \cos{\tilde \beta}=\cos{\frac{\pi}{2}}=0$, and as a result  $ \varphi_{1}^{\text{cl}}$ will not be excited.
 Furthermore, the matrices $t_i$ realise a $k$-dimensional irreducible representation of $\mathfrak{su}\left(2\right)$ satisfying
\begin{IEEEeqnarray}{c}\label{tmatr}
\left[t_i, t_j\right] = i\,\epsilon_{ijl}\,t_l, \qquad i,j,l = 1,2,3.
\end{IEEEeqnarray}
and
\be\label{comm}
t_i=\frac{1}{2}\sum_{j=1}^3\left[t_j,\left[t_j,t_i\right]\right]
\ee
which is fully consistent with \eqref{tmatr}.
The expressions in the equations \eqref{sol-2} and  \eqref{sol-2a}  determining the classical values for the six scalars should be solutions of the equations of motion of  ${\cal N}=4$ SYM 
which are written in \eqref{eoms}. The first thing to notice is that the gauge field $A$ given in \eqref{A-sing} commutes both with itself,  as well as with the vevs of the scalars in \eqref{sol-2} and \eqref{sol-2a}. This means that all covariant derivatives become usual partial derivatives, i. e. $D_\mu=\partial_\mu$ on the solution.

Taking into account the aforementioned comments  the equations of motion \eqref{eoms} simplify to \footnote{The equations of motion should be satisfied everywhere except the surface on which the operator is located.}
\begin{IEEEeqnarray}{c}\label{eoms1}
\left[\partial_{\nu}\varphi_i,\varphi_i\right]=0, \qquad \partial^{\mu}\partial_{\mu}\varphi_i = \left[\varphi_j,\left[\varphi_j,\varphi_i\right]\right].
\end{IEEEeqnarray}
Due to the symmetry of our D5-brane solution we can assume that the scalar fields
depend only on the radial distance from the defect $r'=\sqrt{x'^2_2+x'^2_3}$ and the polar angle $\psi$, namely the coordinates of the plane perpendicular to the surface $\S$. That is we assume that $\varphi_i=\varphi_i(r',\psi)$. In such a case the second equation in \eqref{eoms1} further simplifies to 
\be\label{radial}
\frac{\partial^2\varphi_i}{\partial r'^2}+\frac{1}{r'}\frac{\partial \varphi_i}{\partial r'}+\frac{1}{r'^2}\frac{\partial^2 \varphi_i}{\partial \psi^2}= \left[\varphi_j,\left[\varphi_j,\varphi_i\right]\right].
\ee
Its obvious that given the ansatz \eqref{sol-2}, \eqref{sol-2a} the first equation in \eqref{eoms1} is satisfied.
We now argue that the second equation \eqref{radial} is also satisfied.
Firstly we notice that the scalar fields involving the matrix $C$ solve the homogeneous equation corresponding to   \eqref{radial}, that is, it solves \eqref{radial} with the right hand side being zero. This is enough since all commutators involving $ \varphi_{1}^{\text{cl}} $, $ \varphi_{2}^{\text{cl}} $ and $ \varphi_{3}^{\text{cl}} $ are zero. The only non-zero commutators are those involving only $ \varphi_{4}^{\text{cl}} $, $ \varphi_{5}^{\text{cl}}\ $ and $ \varphi_{6}^{\text{cl}} $. Then, by the use of \eqref{comm},
one can easily show that equation \eqref{radial} is also satisfied for the scalars $ \varphi_{4}^{\text{cl}} $, $ \varphi_{5}^{\text{cl}} $ and $ \varphi_{6}^{\text{cl}}$. In both cases we have taken into account that $[\varphi_i,\varphi_{i+3}]=0\, \,\,\,i=1,2,3$.

We conclude that \eqref{sol-2} and  \eqref{sol-2a} solves the equations of motion for the scalar fields.  Two important comments are in order. Firstly, notice that the solution \eqref{sol-2} and  \eqref{sol-2a}, being the dual description the 
D5 probe brane presented in section \ref{emb_ansatz}, interpolates between the dCFT dual to the half-BPS D3-D3 system\footnote{Actually, as mentioned in the paragraph below equation \eqref{embedding-D3} the gravity dual is a singular D5 brane whose worldvolume and supersymmetry is identical to that of the D3-D3 system. } \cite{Gukov:2006jk}, for the value $\tilde\psi_0=\frac{\pi}{2}$, and the non-supersymmetric dCFT presented in \cite{Georgiou:2025mgg} (see eq. (5.6) of this work)\footnote{To get precise agreement notice that one should rescale the matrices of \cite{Georgiou:2025mgg} as follows $t_i\rightarrow t_i/\sqrt{2}$.}, for $\tilde\psi_0=0$. In particular, for the singular case in which  $\tilde\psi_0=\frac{\pi}{2}$ notice that $k\sim \kappa\rightarrow 0$ and the first equation in \eqref{sol-2}  becomes meaningless, implying that  $ \varphi_{4}^{\text{cl}} $, $ \varphi_{5}^{\text{cl}} $ and $ \varphi_{6}^{\text{cl}}$ should be set to zero in the strict 
$\tilde\psi_0=\frac{\pi}{2}$ limit. At the other endpoint $\tilde\psi_0=0$ we get $ \varphi_{1}^{\text{cl}}=\varphi_{2}^{\text{cl}}=\varphi_{3}^{\text{cl}}=0$
while the other three scalars become the same with the three non-zero scalars of the dCFT of \cite{Georgiou:2025mgg}.

Up to this point, our discussion has  an important caveat.
Notice that, due to the non-periodicity of the the angle $\psi$, the surface operator can  not be solely characterised from the value of the singular gauge field $A$.
Instead, it should be characterised by the expectation value of the Wilson line
\begin{eqnarray}\label{Wilson}
\langle W \rangle &=& \langle Q_l^\dagger(y) (P e^{i \int_{\g_{xy}} A_\mu dx^\mu})Q_l(x)\rangle 
\nonumber\\[5pt]
&=&e^{i\, 2 \pi \rho\, r   \a_1} \sum_{c=k+1}^{N_1+k}\langle Q_{cl}^\dagger Q_{lc} \rangle+  \sum_{c=N_1+k+1}^{N}\langle Q_{cl}^\dagger Q_{lc} \rangle\, ,
\end{eqnarray}
where $\g_{xy}$ is the circular path connecting the initial point $x=(x^0,x^1,r,\phi_0)$ to the end point $y=(x^0,x^1,r,2 \pi \rho+\phi_0)$.\footnote{This Wilson loop crosses itself. It would be interesting to see if it mixes with other configurations under the renormalisation group \cite{Georgiou:2009mp}.}  At this point, let us mention that $\langle Q\rangle$ and $\langle Q^\dagger\rangle$ denote the vevs of the scalars that live on the co-dimension 1 defect which the D7 branes induce (see appendix \ref{Appendix:details_FT-dual} ). Furthermore, because of \eqref{A-sing} and the fact that $Q_{lc}=0,\,\,\, c=1,\ldots, k$ (see comment after \eqref{eoms-3scalars}), $c$ should be summed over $k+1,\cdots,N$ only. This, of course,  holds for the case {\color{blue}(a)} of appendix \ref{Appendix:details_FT-dual}. For the case {\color{blue}(b)}, the first term in the second line of \eqref{Wilson} is absent because in this case $Q_{lc}=0$ for $c=1,\ldots, N_1+k$ and the sum over $c$ should start from $N_1+k+1$. 

This Wilson line is the gauge invariant analogue of the monodromy $\oint A$ in the case of the 1/2-BPS surface operators \cite{Gomis:2007fi}.
Notice here the crucial difference compared to the 1/2-BPS surface operators of \cite{Gomis:2007fi}. We see that the value $a_1$ of the gauge field in \eqref{A-sing} is not enough to fully characterise the operator as is the case for the supersymmetric Gukov-Witten defects. One needs to know the expectation values of the scalars in the hypermultiplets $Q_l$. These fields of the hypermultiplets are necessary in order to define a gauge invariant object, the Wilson line $W$. As mentioned above, all this happens because in our D5 brane solution the angle $\psi$ does not close on itself since generically $\rho\neq 0$. So instead of having a Wilson loop, which is by construction gauge invariant, one has a Wilson line. The restoration of gauge invariance is possible only because of the existence of the doublets $Q_l$ which transform in the fundamental representation of the gauge group $U(N)$. This is in perfect agreement with the gravity side where the gauge invariance of the D5 brane action is only possible if one introduces the D7-branes and puts a magnetic monopole at the boundary of the D5 brane worldvolume (see \eqref{D5-monopole}). As a result, the expectation values of the scalars $Q$ should be related to the charge density of the magnetic monopole at the boundary of D5. It would be interesting to find precise relation between the aforementioned quantities.

As mentioned above, one has the freedom to turn on a Wilson line both for the gauge field $A$ and the dual gauge field $\tilde A$ defined on the D5 probe brane
worldvolume. %along the non-contractible $S^1$.
 As a result, the brane, as well as the dual surface operator, is also characterised by the parameters 
$\eta$.
%In the case of a large number of D5 branes $N_1\gg1$, the probe approximation will no longer be valid and one should find the full back reacted geometry. As it happens for the supersymmetric D3-D3 system, the solution will presumably have a non-trivial disc $D_1$ which ends on the boundary of the asymptotically $AdS_5$ space on a non-contractible
%S^1$ and
%which will be the fiber of $S^1$ over  a straight segment connecting a point in the bulk, at which the $S^1$ shrinks to a point, to a point in the boundary of the asymptotically $AdS_5$ space.
%As a result, in order to fully specify the supergravity solution one should also determine the holonomies of the 2-form gauge fields,  the NS-NS sector 2-form $B_{NS}=dA$
%and  the RR sector 2-form $B_R=d\tilde A$ around the disk. These holonomies are  obviously the gravity counterparts of the field theory parameters $a_1$ and $\eta_1$.
The identification of the parameters at the two sides of the duality is the following 
\begin{IEEEeqnarray}{c}\label{correspondence}
\b_1+i \g_1=\frac{\sqrt{\l}}{2\pi} \frac{1}{\s} e^{i \phi_0}, \quad \eta_1=\oint\frac{\tilde A}{2 \pi} \quad  %a_1=\oint\frac{A}{2 \pi\sin{\tilde\psi_0}}
\end{IEEEeqnarray}
Furthermore, as discussed in the last paragraph the defect is also characterised by the expectation value of the Wilson loop $W$ in \eqref{Wilson} which combines the strength $a_1$ of the gauge field $A$ with the vevs for the hypermultiplets' scalars $\langle Q_{cl}^\dagger Q_{lc} \rangle$.

Some final comments are in order. In the dual gravity description, the dimension of the representation $k$ appearing in \eqref{sol-2} and   \eqref{sol-2a} corresponds to the integer flux through the $S^2\subset S^5$ (see \eqref{kappa}). Let us mention that, in contradistinction to other defect CFT setups,   $k$ cannot be zero in our construction, the mere existence of our solution generically requires $k\neq 0$.
At this point, let us make some further comments regarding the symmetries of 
the field theory solution of the current section. As can be seen from \eqref{sol-2} and   \eqref{sol-2a}  the symmetry of the solution is $SO(2,2)\times SO(2)\times SO(3)$. This symmetry perfectly matches the $AdS_3\times S^1\times S^2$ symmetry of the D5 probe brane. In particular, the isometries of the $AdS_3$ space correspond to the conformal group in the $1+1$ dimensions of the 
surface defect.
Also note that in our solution, as in the case of the co-dimension 2 supersymmetric D3-D3 system  \cite{Drukker:2008wr},  there is a relation between the angle  $\psi$ and the internal angle $\tilde \gamma$, namely ${\psi} -\rho \,\tilde \gamma =\phi_0$. The angle $\tilde\g\in [0,2 \pi)$ corresponds to the $S^1$ isometry of our solution.
 Finally, notice that, 
on the field theory side, the conformal symmetry of the duality is manifest in the form of the solution \eqref{sol-2} or  \eqref{sol-2a} that scales as $\frac{1}{r'}$ and $\frac{1}{z}$. %Finally, the choice of the above solution is dictated by the fact that in the dual gravity theory the D5-brane sits at $\tilde \psi=0$ which implies that only the three scalars related to one of the $S^2$s of \eqref{metric} should have non-zero vacuum expectation values (vevs), namely $\varphi_{i+3}, \,i =1,2,3$.

Before closing this section let us make a final important comment. Although we have shown that the D5 brane has no tachyonic excitations since the masses of all the fluctuations are above the B-F bound we have not proved that the full system of the two D7 and the D5 brane sits at an equilibrium point. One might worry that the D7 branes will attract or repel each other since we are dealing with a non-supersymmetric configuration. This interaction will take place by closed string states emitted by one of the D7 branes and absorbed by the other. On one hand, the corresponding cylinder diagram will be proportional to $g_s^0$, where 
 $g_s\sim \frac{1}{N}$ is the string coupling. 
On the other hand, the mass of the D7 branes will be proportional to its tension which scales with the inverse $g_s$, that is $M_{D7}\sim T_7\sim \frac{1}{g_s}$. Due to the fact that we are working in the large $N\rightarrow \infty$ limit the D7 brane becomes extremely heavy and thus imperturbable since it acceleration will be $\a\sim g_s\rightarrow 0 $. Another source of instability could be the tension of the D5 brane which pulls the two D7 branes together. In this case one can argue in a similar manner.
This force will be proportional to the tension of the D5 brane which is 
$F\sim T_5\sim \frac{1}{\a'^3}$. However, the mass of the D7 brane scales with $\a'$ like 
$M_{D7}\sim T_{D7} \sim\frac{1}{\a'^4}$ producing an acceleration for the D7 which is $\a=\frac{F}{N_{D7}M_{D7}}\sim \frac{\a'}{N_{D7}}\rightarrow 0 $, where $N_{D7}$ is the number of D7 branes in each stack. Thus, in the worst case, one can argue that one is working within the adiabatic approximation similar to what happens in the Born-Oppenheimer approximation. As long as one calculates correlation functions whose time and length scales are much smaller than the scale set by the acceleration of the D7 branes the latter can be considered as static.

\section{Conclusions}\label{concl}
In this work, we have introduced a new class of holographic dualities between certain non-supersymmetric defect conformal field theories (dCFTs) and their gravity duals. The theories are parametrised by two continuous parameters $\rho$ and $\s$, as well as by monodromies along certain cycles \eqref{eta-sing}, \eqref{Wilson}. On the gravity side, the defect is realised by a  
probe D5 brane wrapping an $S^1\times S^2$ subspace of the internal $S^5$. The symmetry of the induced metric on the D5 brane is $AdS_3\times S^1 \times S^2$.  
The intersection of $AdS_3\times S^1$ with the boundary of $AdS_5$ creates a surface defect whose co-dimension is 2. 
%As mentioned above, the solution depends on two continuous parameters. 
The continuous parameter $\s$ determines the inclination angle of the D5 brane with respect to the boundary of the spacetime, while 
$\rho$ is related to the $S^1$ and determines the winding of the brane around one of the angles of the internal space $S^5$.

An important feature of our construction is that it classically interpolates between the supersymmetric D3-D3 system at one end, and the non-supersymmetric D3-D5 system  presented in \cite{Georgiou:2025mgg} at the other.
Next we examined the stability of our brane configuration. In particular, we determined the regions of the parametric space $(\rho,\s)$
in which the masses of all the fluctuations of the transverse to the D5 brane coordinates respect the B-F bound. In section \ref{D5-D7}, we argued that for our construction to be valid one needs to insert in the geometry two D7 branes on which the D5 brane solution terminates. This is necessary because the "winding" number $\rho$ is not an integer and as a result the D5 brane has boundaries. These boundaries induce a gauge anomaly for the D5 brane which cancels through anomaly inflow from the D7 branes.
%Hence, the full system of the D5 and D7 branes is {\color{red}stable} and anomaly free.
Hence, the full system of the D5 and D7 branes is
anomaly free and has no tachyonic instabilities for a certain range of its parameters. 

In section \ref{dual}, we discussed the field theory dual CFTs of the D5-D7 system. In particular, we determined the classical solution of the ${\cal N}=4$ SYM equations of motion which we conjecture to describe the co-dimension 2 surface operators and comment on the quantities needed to describe these non-supersymmetric defects. The field theory dual depends, of course, on the same number of parameters as the probe D5-D7 system.  We have also argued that  the effect of the two D7 branes is minimal, since the flux through one of the two-spheres they wrap is zero, and as a result they do not introduce any vevs for the six bulk scalar fields. However, the D7 branes play a crucial role in the cancellation of the gauge anomalies. Furthermore, they are important in ensuring the consistency of the construction since they introduce expectation values for the scalar fields belonging in the "would be" hypermultiplets which live on the defect. These expectation values are necessary because otherwise one would not be able to define a gauge invariant Wilson line (see eq. \eqref{Wilson}) given the fact that in the dual description the boundary angle $\psi$ is not periodic. The presence of the aforementioned expectation values at the end of the Wilson line restores the gauge invariance which would be otherwise violated. In the dual string theory picture the role of the D7 branes, that are the source for the expectation values of the scalars $\langle Q\rangle\neq 0$, is also crucial because they cancel the gauge anomaly of the D5 brane through anomaly inflow and by the introduction of a magnetic monopole supported on the 5-dimensional  boundary $\partial M_{D5}\subset M_{D7}$ of the D5 brane. 

Our work opens several directions for future research. First of all, one can calculate the anomaly coefficients of the co-dimension 2 defects introduced in this paper.  This may be done at strong coupling by exploiting the gravity solution, as well as at weak coupling by using the conformal field theory defect CFT. Furthermore, in a similar manner, one can  calculate one-point correlation functions of the stress-energy tensor and of chiral primary operators at both weak and strong coupling. One can then compare the results for the aformentioned observables and check if, in the appropriate limit, the two results agree, as it occurred in \cite{Georgiou:2025mgg}. This will be strong evidence that the field theory dual of section \ref{dual} is complete. This is not completely obvious because the low amount of symmetry does not constrain the field theory as much as it does for the co-dimension 2 supersymmetric Gukov-Witten defects.
Another important direction would be to calculate holographically higher point correlation functions of operators belonging to protected multiplets \cite{Georgiou:2023yak}. Furthermore, it would be interesting to specify how the parameters characterising the defect transform under S-duality.
Finally, one may calculate correlation functions of operators located on the defect. This could help one to clarify the details of the dynamics of the degrees of freedom living on the defect, that is to fully determine the precise form of $S_{def}$.

\subsection*{Acknowledgements}

This paper has been financed by the funding programme ``MEDICUS", of the University 
of Patras (D. Z. with grant number: 83800).

%\newpage
\appendix\section[Supersymmetry]{Supersymmetry \label{Appendix:Supersymmetry}}
In this Appendix,  we study the Poincare and conformal supersymmetries  preserved by  the solution given in \eqref{sol-2}. These symmetries are generated by two ten dimensional Majorana-Weyl spinors  $\epsilon_1$ and  $\epsilon_2$ of opposite chirality.
To determine the  supersymmetries  that are left unbroken by \eqref{sol-2}  
we study the supersymmetry variation of the gaugino. The number of independent spinors for which the supersymmetry variation of the gaugino is zero gives the number of the preserved superconformal symmetries.

The metric of the CFT is:
\noindent
%$\bullet$
\be\label{minkowskico}
ds^2=-(dx^0)^2+(dx^1)^2+(dx^2)^2+(dx^3)^2=-(dx^0)^2+(dx^1)^2+dr^2+r^2 d\psi^2  .
\ee
A Poincare supersymmetry variation is  given by 
%\eqnn\susygaugef\eqnn\susygaugeff
\begin{eqnarray}\label{susyvar}
\delta \psi= \Big(\frac{1}{2}F_{\mu\nu}\Gamma ^{\mu\nu}+D_{\mu}\varphi_i\Gamma^{\mu\, i+3}-\frac{i}{2}[\varphi_{i},\varphi_{j}]\Gamma^{i+3\,\,j+3}\Big) \epsilon_{1}% .
\end{eqnarray}
In \eqref{susyvar} $\mu$ runs from 0 to 3 while $i$ runs from 1 to 6. 
As usual, $\Gamma ^{\mu}$ and $\Gamma ^{i+3}$ denote the 10-dimensional gamma matrices satisfying 
$\{\Gamma_M,\Gamma_N\}=-2 \,\eta_{MN}{\mathbf 1}_{10}$, where the indices $M,N=0,1,\dots,9$ and $\eta_{MN}$ is the mostly plus flat metric in 10 dimensions. By inserting \eqref{sol-2} in \eqref{susyvar} one obtains
\begin{eqnarray}\label{supervar-2}
\delta \psi=-\frac{x'_2}{\sqrt{2}r'^3}\,\Gamma^{2, i+6}\epsilon_{1}\otimes {\cal T}_i -\frac{x'_3}{\sqrt{2}r'^3}\,\Gamma^{3, i+6}\epsilon_{1}\otimes{\cal T}_i  +\frac{1}{4 \,r'^2}\,\epsilon_{ijl}\, \Gamma^{i+6,j+6}\epsilon_{1}\otimes{\cal T}_l+\ldots,
\end{eqnarray}
where the dots stand for terms proportional to $\Gamma^{2, i+3}$ and $\Gamma^{3, i+3}$ which have different spacetime dependence compared to the term written explicitly, and as a result they will not intervene in our argument. Furthermore,  $i=1,2,3$ and ${\cal T}_i=\left[\begin{array}{cc} \left(t_i\right)_{k\times k} & 0_{k\times \left(N - k\right)} \\ 0_{\left(N - k\right)\times k} & 0_{\left(N - k\right)\times \left(N - k\right)} \end{array}\right]_{N\times N}$. 
Now if we focus on the term proportional to $\frac{x'^2}{r'^3}$ we get
\be
\delta \psi=0\quad  \Rightarrow \quad \Gamma^{2,i+6}\epsilon_{1}=0 \quad  \Rightarrow \quad
\epsilon_{1}=0
\ee
because the matrix $\Gamma^{2,i+6}$ is invertible. We conclude that none of the 16 supersymmetries of the $AdS_5\times S^5$ background is preserved in the presence of the defect.

In a similar manner, the superconformal supersymmetry transformations are given by
%\eqnn\sucongaugef\eqnn\sucongaugeff
\begin{eqnarray}\label{supercon}
\delta \psi&=\Big( (\frac{1}{2}F_{\mu\nu}\Gamma ^{\mu\nu}+D_{\mu}\varphi_i\Gamma^{\mu\, i+3}-\frac{i}{2}[\varphi_{i},\varphi_{j}]\Gamma^{i+3\,\,j+3}) x^{\sigma}\Gamma _{\sigma}-2\varphi_{i}\Gamma^{i+3}\Big) \epsilon_{2}
\end{eqnarray}
and by focusing on the term originating from the last term in \eqref{supercon}
we get
\be
\delta \psi=0\quad  \Rightarrow \quad \Gamma^{i+6}\epsilon_{2}=0 \quad  \Rightarrow \quad
\epsilon_{2}=0, \, \quad \textrm{with } \quad i=1,2,3.
\ee
In agreement to the strong coupling analysis, our final and rather obvious conclusion is that our system is non-supersymmetric.
In the limit $\tilde\psi_0=\frac{\pi}{2}\Rightarrow k=0$, and in agreement to the $\kappa$-symmetry analysis of section \ref{interpol}, the field theory solution becomes 1/2-BPS. This is because the contribution written explicitly in \eqref{supervar-2} is zero while the remaining part denoted by the dots gives the same 1/2-BPS conditions, since the vevs of the scalars become those dual to the D3-D3 supersymmetric system \cite{Gomis:2007fi}.

%%%%%%%%%%%%%%%%%%%%%%%%%%%%%%%%%%%%%%%%%%%%%%%%%%%%%%%%%%%%%%%%%%%%%%%%

\section{Details regarding the magnetic monopoles}
\label{Appendix:details_monopoles}
The careful reader may have noticed that in the last term of \eqref{D5-C4} the value of the worldvolume 2-form that we have used is that of the D5 solution.
However, there is a subtle point. In order to cancel the anomaly we have inserted two magnetic monopoles, one monopole and one anti-monopole, sitting at the boundaries of the D5 brane, one at $\psi=\phi_0$ and the other at 
$\psi=2\pi \rho +\phi_0$. These monopoles create an additional contribution in the worldvolume field $F$ which can potentially  contribute in the right hand side of \eqref{D5-C4}. Here we will argue that this will not happen. Firstly notice that \eqref{D5-monopole} can be easily integrated to give $F_{monopole}\sim \theta(\psi=\phi_0)\, d \b\wedge d\g$. But the integral of the last term of  \eqref{D5-C4} is performed on $\partial M_{D5}$ on which the argument of the theta function is zero, and as a result its value is ill-defined. To resolve this issue we place the monopoles not exactly on the boundary but a small distance $\epsilon$ from the boundary, i. e. inside the D5 brane worldvolume, namely at $\psi=\phi_0+\epsilon$ and $\psi=2\pi \l+\phi_0-\epsilon$. At 
the end of the calculation one should send, of course, $\epsilon \rightarrow 0$. In particular, we have
\begin{IEEEeqnarray}{l}\label{two-mono}
dF_1^{(D5)}=2 \pi\, g_{D5}\,\sin\b\,\d(\psi-(\phi_0+\epsilon))\,  d\psi \wedge d \b \wedge d\g \nonumber \\[5pt]
dF_2^{(D5)}=-2 \pi\, g_{D5}\,\sin\b\,\d(\psi-(2\pi \l+\phi_0-\epsilon))\,  d\psi \wedge d \b \wedge d\g\, .
\end{IEEEeqnarray}

The next step is to integrate  \eqref{two-mono} to get the worldvolume 2-form generated by the two magnetic monopoles. We obtain at the two ends
\begin{IEEEeqnarray}{l}\label{two-mono-1}
F_1^{(D5)}= 2 \pi g_{D5}\sin\b\,\theta (\psi-\phi_0-\epsilon)\, d\b\wedge d\g\nonumber \\[5pt]
F_2^{(D5)}= 2 \pi g_{D5}\sin\b\,\, \theta (2\pi \l+\phi_0-\epsilon-\psi)\, d\b\wedge d\g\, .
\end{IEEEeqnarray}
From \eqref{two-mono-1} we see that on the boundary of D5, that is precisely at  $\psi=\phi_0$ and $\psi=2\pi \rho+\phi_0$ the theta functions above vanish since their argument is negative, $\theta(-\epsilon)=0$. We conclude that the contribution of the monopoles in the last term of \eqref{D5-C4} can be safely ignored.

\section{Some more details of the field theory dual }
\label{Appendix:details_FT-dual}
%%%%%%%%%%%%%%%%%%%%%%%%%%%%%%%%%%%%%%%%%%%%%%%%%%%%%%%%%%%%%%%
One may be puzzled since it seems that, for the case of the D7 branes with one of the fluxes being zero, there is no difference compared to the case of ${\cal N}=4$ SYM without any defect since the vevs of all scalars are set to zero. The resolution of this apparent puzzle is simple. The complete action is that of the bulk field theory plus the action governing the dynamics of the fields localised on the defect and is schematically given by \cite{DeWolfe:2001pq}
\begin{IEEEeqnarray}{ll}
S_{tot}=S_{\N = 4} +S_{def}, \quad {\rm where} \quad 
\nonumber \\
S_{def}=-\frac{1}{g_{YM}^2}\int_{\psi=\psi_0} d^3x\Bigg (
\,(Q_{c_1l}^\dagger \,\s_i\, T^a_{c_1c_2} Q_{l c_2}) \,\,\frac{1}{r}\partial_\psi\tilde\varphi^a_i|_{\psi=\psi_0}+(Q_{c_1l}^\dagger  T^a_{c_1c_3} T^b_{c_3c_2}Q_{l c_2}) \hat \varphi_i^a\hat \varphi_i^b
\nonumber\\
\qquad \quad \, -\frac{1}{2}\varepsilon_{ijk}f_{abc}\,(Q_{c_1l}^\dagger \,\s_i\, T^a_{c_1c_2} Q_{l c_2})\tilde\varphi_j^b \tilde\varphi_k^c+\mathcal{O}(Q^4)\,
\Bigg )+S_{kin}+S_{Yukawa} .
\label{defect-Lag}
\end{IEEEeqnarray}
In \eqref{defect-Lag}, $Q_l=\left(\begin{array}{cc} q_{l\,c}  \\ \tilde q^\dagger_{l\,c} \end{array}\right), \,\,\, l=1,\cdots k_{D7},\, c=1,\cdots N$ denotes the $2$ complex scalar fields belonging to each of the $k_{D7}$ "would be" hypermultiplets that live on the 3-dimensional co-dimension 1 defect created by the D7 branes.\footnote{As usual, 
$2 \pi k_{D7}=\int_{S^2} F^{(D7)}$, with $F^{(D7)}$ being the worldvolume gauge field obtained from the D7 solution \eqref{sol-D7}.} These hypermultiplets correspond to strings whose one end is located on the D7 branes while the other on the D3 branes generating the $AdS_5\times S^5$ geometry. Consequently, $q$ transforms in the $(k_{D7}, N)$ representation of the $U(k_{D7})\times U(N)$ group while $\tilde q$ transforms in the $(\bar k_{D7}, \bar N)$ representation of the $U(k_{D7})\times U(N)$ group. Thus, $q$ can be viewed as a $k_{D7}\times N$ matrix while $\tilde q$ as an $N\times k_{D7}$ matrix and the moment map $(\mu_i)_{c_1 c_2}=Q_{c_1 l}^\dagger \s_i Q_{l c_2}$, or equivalently $(\mu_i)^a=Q_{c_1 l}^\dagger \s_iT^a_{c1c2} Q_{l c_2}$, is an $N\times N$ matrix.\footnote{We remind the reader that $N$ is the number of colours.} At this point, we should stress that the fields $q$ and $\tilde q$ live on the defect and one way they interact with the bulk is through the bulk scalars $\varphi_i$ with the latter being localise on the 3-dimensional defect.

Here we distinguish two cases: {\color{blue}(a)} when the non-zero flux of the D7 is through the $S^2$ parametrised by $(\tilde \b,\tilde\g)$. In this case,
the bulk scalar fields $\tilde\varphi_i,\, i=1,2,3 $ appearing in the first and last term of \eqref{defect-Lag} are $\varphi_i,\, i=1,2,3 $ while those appearing in the second term are  $\hat\varphi_i=\varphi_{i+3},\, i=1,2,3 $, and {\color{blue}(b)} when the non-zero flux of the D7 is through the $S^2$ parametrised by $( \b,\g)$. In this case, the role of the two sets of scalar fields is exchanged, that is
$\tilde\varphi_i=\varphi_{i+3},\, i=1,2,3 $ appear in the first and last term of \eqref{defect-Lag} while $\hat\varphi_i=\varphi_{i},\, i=1,2,3 $ appear in the second term (see eq. \eqref{coord-sphere}).

%Furthermore, 
%when the non-zero flux of the D7 is through the $S^2$ parametrised by %$(\tilde \b,\tilde\g)$,
%the bulk scalar fields appearing in the first and last term of \eqref{defect-Lag} are $\varphi_i,\, i=1,2,3 $ while those appearing in the second term are  $\hat\varphi_i=\varphi_{i+3},\, i=1,2,3 $ . In the case where the non-zero flux of the D7 is through the $S^2$ parametrised by $( \b,\g)$, the role of the two sets of scalar fields is exchanged.
%$\varphi_{i+3},\, i=1,2,3 $ appear in the first and last term of \eqref{defect-Lag} while $\hat\varphi_i=\varphi_{i},\, i=1,2,3 $ appear in the second term (see eq. \eqref{coord-sphere}).

Using now \eqref{defect-Lag} the equations of motion for the three plus three scalars become \footnote{In both equations of \eqref{eoms-3scalars} the sum involving the index $j$ is over all six scalars, i. e. $j=1,\ldots, 6$.}
\begin{IEEEeqnarray}{c}\label{eoms-3scalars}
D^{\mu}D_{\mu}\tilde\varphi_i^a - \left[\varphi_j,\left[\varphi_j,\tilde\varphi_i\right]\right]^a=- (\mu_i)^a  \,\frac{1}{r}\partial_\psi\d(\psi-\psi_0)-\varepsilon_{ijk}f_{abc}(\mu_j)^b \tilde\varphi_k^c \, \d(\psi-\psi_0)\nonumber\\
D^{\mu}D_{\mu}\hat\varphi_i^a - \left[\varphi_j,\left[\varphi_j,\hat\varphi_i\right]\right]^a=Q_{c_1l}^\dagger  \{T^a, \hat\varphi_i\}_{c_1c_2}Q_{l c2} = 0. %\\
\end{IEEEeqnarray}
The second equation above gives zero because one can always make the choice $\hat\varphi_i Q=0=Q^\dagger\hat\varphi_i $. Given the structure of the expectation values of the six scalars induced by the D5 branes \eqref{sol-2} this can always be achieved. Indeed, in the case {\color{blue}(a)} mentioned above 
one can choose the first $k$  entries of each $Q_l$ to be zero, namely that $Q_{l c}=0$ when $c=1,\ldots, k$ \cite{Arean:2006vg}.
In the case {\color{blue}(b)} the equation $\hat\varphi_i Q=0=Q^\dagger\hat\varphi_i$ is satisfied if one chooses only the first $N_1+k$  entries of each $Q_l$ to be zero, namely that $Q_{l c}=0$ when $c=1,\ldots, N_1+k$. Here $N_1$ is the number of the D5 branes.
Notice now that when $\varphi_i=0=\hat\varphi_i$, as it happens for our D7 brane, the first equation is satisfied when $\mu_i=0$.\footnote{Even when the vevs of the 6 scalars are given by the D5 profile \eqref{sol-2}, which is the case in the presence of the D5 brane, the left hand side of \eqref{eoms-3scalars} is still zero implying that $\mu_i=0$. } This last equation does not necessarily imply that the expectation values for the scalars of the hypermultiplets is zero. In fact, one can reside at the Higgs branch in which $\langle q \rangle\neq 0$ and $ \langle \tilde q \rangle\neq 0$.
%%%%%%%%%

%\section{Details of the computations}
\label{Appendix:details_computations}

\bibliographystyle{utphys}

\bibliography{refs}

\end{document}